\documentclass[aps,prfluids,preprint,superscriptaddress]{revtex4-2}
\usepackage{graphicx}
\usepackage[scriptsize]{subfigure}
\usepackage{IEEEtrantools}
\usepackage{amsmath,amssymb}
\usepackage{hyperref}
\newcommand{\sgn}{\mathrm{sgn}}
\newcommand{\ud}{\,\mathrm{d}}
\renewcommand{\i}{\mathrm{i}}
\newcommand{\figwidth}{15.2cm}
\newcommand{\e}{\mathrm{e}}

\begin{document}


\title{Analysis and Prediction of Noise from Installed Subsonic Chevron Jets}



\author{Hasan Kamliya Jawahar}
\affiliation{Department of Aerospace Engineering, University of Bristol, Bristol BS8 1TR, UK}

\author{Benshuai Lyu}
\email[Corresponding author:]{b.lyu@pku.edu.cn}
\affiliation{School of Mechanics and Engineering Science, Peking University, Beijing 100871, China}

\author{Mahdi Azarpeyvand}
\affiliation{Department of Aerospace Engineering, University of Bristol, Bristol BS8 1TR, UK}



\begin{abstract}
An experiment is conducted to investigate the effects of chevrons on installed subsonic jet noise at a Mach number of $0.5$ using the NASA SMC000 (round) and SMC006 (chevron) nozzles. The jets are of a diameter $D=16.93\mathrm{mm}$ and placed near a flat plate, with a horizontal separation distance $L=6.5D$ between the plate's trailing edge and the nozzle exit. The vertical separation distance $H$ varies between $1.5D$, $2D$, $2.5D$ and $3D$. Far-field sound is measured at various observable angles ranging from $\theta=60^\circ$ to $150^\circ$ to the downstream jet axis on the reflected side. The measured sound spectra are compared to the near-field scattering model developed by Lyu et al, together with the isolated near-field pressure spectra inputs from corresponding Large Eddy Simulations (LES) for both nozzles. Results show that jet installation results in a strong noise amplification at low frequencies for both nozzles due to the scattering of the near-field pressure fluctuations and a mild noise increase at high frequencies due to surface reflection (reflected side). The low-frequency amplification is strongest at $H=1.5D$ and has a dipolar directivity. A secondary spectral hump appears within this low-frequency amplification hump, which is hypothesised due to the interference between the sound generated by the large coherent structures directly and that by their scattering at the plate's trailing-edge. The use of chevrons reduces the low-frequency noise for isolated jets, but leads to even stronger noise amplification for installed jets; this is likely due to enhanced jet mixing resulting in stronger near-field pressure fluctuations at a fixed radial distance. Results show that the scattering model can predict the low-frequency noise amplification well at various observer angles for both nozzles, suggesting the validity of the instability-wave scattering mechanism and modelling for both round and chevron jets.
\end{abstract}
\newpage


\maketitle
\newpage




\section{Introduction} \label{sec:Intro}

Due to the rapid expansion of the global passenger aircraft fleet, aircraft noise has become a critical concern for both aircraft manufacturers and airport authorities. The persistent and loud noise generated by aircraft can lead to various negative consequences for individuals residing near airports, including hearing impairment, sleep disturbance, and decreased cognitive functioning. As such, increasingly stringent noise regulations are put into place as an important requirement during aircraft certification.

Aircraft noise typically includes multiple components. Among all the noise sources, jet noise remains the dominant source during takeoff. 
Extensive research has been conducted to develop technologies that reduce jet noise. 
In the mid-1990s, the addition of chevrons - a sawtooth-like pattern - to the trailing edge of exhaust nozzles marked a significant development in high-speed jet aircraft research. The specific aim of this innovation was to reduce the jet noise produced by these aircraft, particularly during takeoff and landing. The term ``chevron” seemed to first appear in studies regarding mixer-ejector nozzles under the HSR (High Speed Research) program and then in the 1995 GEAE (General Electric Aircraft Engines) proposal to the AST (Advanced Subsonic Technology) Program of NASA~\cite{zaman2011evolution}.

Various experimental studies have confirmed that chevrons possess noise reduction capabilities, and these effects have been attributed to several underlying mechanisms. Chevrons can reduce low-frequency jet noise \cite{gudmundsson2007spatial,bridges2004parametric,kamliya2021effects}, which is commonly associated with large-scale structures in the flow, and increase high-frequency noise, usually attributed to small-scale turbulent structures \cite{gudmundsson2007spatial}. Krasheninnikov and Mironov showed in their study that chevrons reduce noise by producing circulation in the secondary flow in a plane normal to the jet axis. The streamwise vorticity component thus generated enhances mixing, leading to diminished jet noise. This also results in a reduction in turbulent fluctuation energy (in accordance with the Townsend model) \cite{krasheninnikov2003effect,krasheninnikov2003turbulent}. Induction of vorticity in the shear layer has also been shown to shorten the potential core length. Moreover, it has been observed that the sources of jet noise shift closer to the nozzle due to the impact of chevrons \cite{nikam2014aero}. Thus noise reduction mechanism essentially involves shifting acoustic energy to higher frequencies, whose more effective attenuation by the atmosphere may also be of help.

Similarly, Saiyed et al.~\cite{saiyed2000acoustics} conducted experiments to assess the acoustics and thrust of high BPR engines and found that a noise reduction of up to 2.7~dB EPNL can be achieved using chevrons in comparison to round nozzles. Moreover, designs in which both the core and fan nozzles were modified showed the greatest EPNL reductions after correcting take-off thrust loss.

Nikam and Sharma~\cite{nikam2017effect} investigated the effects of the penetration and position of the chevron on the mixing layer in isolated jet configurations. Different levels of chevron penetration were tested at varying longitudinal locations, and the results showed a similar mixing increase, a reduction of the potential core length, and an upstream shift in the dominant noise source. Overall noise levels were higher close to the chevrons, but were reduced further downstream compared to the round nozzle. Noise from the chevron nozzle was also directed towards higher polar angles at higher frequencies.
Similar noise reduction was observed in another study conducted on separate-flow nozzles by Mengle et al. \cite{mengle2006reducing}. They used different types of azimuthally varying chevrons (AVC) capable of modulating the mixing around the nozzle periphery. It was revealed that certain combinations of modified fan and core AVC nozzles were quieter than the conventional chevron nozzles or baseline (round) nozzles. 

In 2004, Bridges and Brown~\cite{bridges2004parametric} tested a series of chevron nozzles with varying chevron count, penetration, length, etc, in isolated jets in order to uncover the relationships between these parameters and flow/noise characteristics. They concluded that chevron penetration increases noise at high frequencies while reducing the noise at low frequencies. In particular, with a large number of chevron teeth, a good reduction in low-frequency noise can be achieved due to enhanced mixing in the shear layer. 

Most of the investigations regarding the impact of chevrons on the flow and acoustic fields have been conducted on isolated jets. Recent years, however, have witnessed a new important issue concerning the noise emitted by the high-speed jets from aircraft, i.e. with the introduction of high-bypass-ratio engines, jet noise is significantly enhanced at low frequencies due to the interaction between the jet flow and nearby wings and pylons~\cite {Mead1998,hunter2005computational,elkoby2005full,Cavalieri2014,Lyu2017a}.
This is typically referred to as jet installation effects, and the noise generated in the installation configurations is often referred to as installed jet noise.

Due to its significant technological relevance, recent research has started to investigate the possibility of using chevron nozzles to reduce installed jet noise. The effect of chevrons on the jet installation noise has been explored in several experimental studies. To investigate the potential of jet noise shielding in a Hybrid Wing Body (HWB) airplane, chevrons were introduced as one of the devices to redistribute the jet noise source~\cite{mayoral2010effects}. Both mild and aggressive chevron configurations were applied with and without the combination of porous wedge fan flow deflectors, and a flat plate in the shape of the HWB planform was used to incur the shielding effect. It was found that the application of aggressive chevrons yielded a noise reduction of 6.5~dB EPNL, and when they were combined with fan flow deflectors, the noise benefit increased to 7.6~dB. The observed noise reduction was achieved due to the significant reduction in the potential core length.

Mengle~\cite{mengle2011effect} experimentally studied installed jet noise near a model-scale wing with trailing-edge flaps and fuselage under take-off conditions. The engine had a bypass ratio of 7. The effects of reducing the gulley height, increasing the flap angle, and applying azimuthally varying T-fan chevrons were studied. Smaller gulley heights and larger flap angles generally lead to higher installed noise, but chevrons reduce jet-flap interaction noise. However, chevrons show slightly more scrubbing due to enhanced mixing. Nevertheless, they can be implemented to achieve quieter aircraft if advanced chevron designs are used \cite{mengle2011effect}.

In a subsequent study conducted by Mengle et al.\cite{mengle2007flaperon}, the impact of sawtooth trailing edges and mini-vortex generators on noise caused by jet-flap interaction was investigated for round coaxial and chevron nozzles under both take-off and approach conditions. Results revealed that during approach conditions, the sawtooth trailing edge of the baseline round nozzle had a marginal impact on far-field noise reduction. However, co-rotating mini-vortex generators displayed a slightly superior advantage, reducing noise levels by approximately 1~dB across a wider angular range covering both front and aft quadrants. In contrast, for the advanced RT-chevron nozzle, the co-rotating mini-vortex generators led to negative consequences, creating undesirable jet-flap interaction and failing to provide a beneficial outcome.

Another study regarding jet-wing interaction was performed by Kopiev et al. \cite{kopiev2013intensification} in which they carried out theoretical and experimental investigations of jet noise generated due to interaction between a turbulent jet and a wing. Tests were carried out for different types of nozzle edges under various configurations. A semi-infinite plate was used as the model wing. The results of the experiments without chevrons showed a total increase in noise of 4-5~dB at high frequencies and around 10~dB at low frequencies, depending on the observer angle. However, chevrons at the secondary nozzle resulted in a decrease in noise at low frequencies and an increase at high frequencies. To explain this noise reduction, Kopiev et al. hypothesised that chevrons decrease the amplitude of low-frequency axisymmetrical instability waves that are responsible for low-frequency pseudo-sound in the near field of the jet and for low-frequency noise radiated in the far field. Thus, the decrease in low-frequency pressure fluctuations near the sharp edge results in the decrease of the low-frequency component of scattered sound at the sharp edge.

To study the impact of chevrons on the jet-surface interaction noise, Bastos et al. \cite{bastos2017experimental} performed an experimental investigation involving a cold subsonic jet installed with a flat plate. The far-field noise produced from baseline and chevron nozzles was assessed for different relative positions between the nozzle and the plate. The comparative results for both nozzles in isolated and installed configurations showed that the chevrons were very effective in mitigating the jet-surface noise. However, for highly integrated configurations, the noise reduction by chevrons was almost negligible.

Jawahar and Azarpeyvand~\cite{kamliya2022Hydro} carried out a thorough examination of the hydrodynamic and acoustic characteristics of subsonic jets with chevron nozzles in a recent study. Their primary objective was to determine the sources of noise in unheated jets and establish efficient techniques for reducing noise by analysing the influence of jet installation factors.

Several experiments have been conducted to understand the noise generation mechanism in installed subsonic jets by using chevrons. For instance, Carbini et al.~\cite{carbini2022experimental} investigated the physics of jet-surface interaction noise and showed that the chevron jet resulted in a significant broadband noise reduction compared to the round jet. This reduction in noise was also supported by a decrease in the spanwise correlation length along the wing trailing edge across a wide range of frequencies. Additionally, a significant reduction in the tonal trapped wave energy was observed. In another study, Carbini et al.~\cite{carbini2023experimental} performed an experimental investigation to clarify the physics behind jet-surface interaction and jet mixing noise. Various measurements were taken on a subsonic jet and an airfoil with different nozzle shapes and vortex generator geometries. The findings in this study may be used to improve the design of aircraft engines and reduce their noise emissions.

Besides using chevrons, lobes were also used as a way of reducing installed jet noise. Using this approach, Lyu and Dowling~\cite{Lyu2017b,lyu2019experimental} conducted an experimental study to investigate the effects of lobed nozzles on installed jet noise. They found that the lobed nozzles led to a reduction in installation noise in the intermediate- and high-frequency regimes on the shielded side due to enhanced jet mixing and effective shielding, but little reduction was observed for low-frequency noise due to the scattering of jet instability waves, which is possibly due to the fact that lobed nozzles cause negligible change to the dominant mode 0 (axisymmetric) jet instability waves. On the reflected side of the plate, noise reduction was only observed in the intermediate frequency range, with a slight increase in high-frequency noise due to enhanced jet mixing. Overall, the use of lobed nozzles had some noise benefits in certain frequency regimes, but their effects depended upon the plate position and frequency of the noise.

Besides experiments, recent advancements in computational fluid dynamics (CFD) have played a significant role in studying jet aerocoustics and its installation effects. Thomas et al.~\cite{thomas2001computational} conducted a computational study to investigate the interaction between the pylon and exhaust flow from a dual-stream nozzle with chevrons. The study focused on analysing the impact of the pylon on the symmetrical lobed pattern created by the core chevrons in the jet plume. Computational analysis was carried out by running cases both with and without the pylon, and with the chevron nozzle rotated relative to the pylon. The study found that the pylon's presence distorted the lobed pattern, leading to changes in the generated noise.

In the study conducted by Wang et al.~\cite{wang2017rans}, the impact of chevrons on noise reduction was investigated using a Hybrid Large Eddy Simulation (LES) and Reynolds-Averaged Navier-Stokes (RANS) approach, with and without a finite span wing installation. The findings of this simulation demonstrated that chevrons effectively reduced noise levels in both isolated and installed jets by mitigating the noise sources in the shear layer and the interaction between the jet and the wing/flap. The study also showed that installing chevrons on the outer nozzle of a coaxial jet was an effective way of reducing jet noise, and their performance can be further enhanced by incorporating a finite span wing.

Jawahar et al.~\cite{kamliya2021effects} performed numerical investigations to study the impact of chevrons on the jet-installation noise involving subsonic cold jets. Tests were also experimentally conducted for four different types of chevron nozzles and various plate positions for validation purposes. The characteristics of the jet hydrodynamic pressure fluctuations were investigated features were studied using the Wall-Modelled LES for both isolated and installed configurations. The OASPL showed that chevron nozzles were effective in reducing the installation noise at low Mach numbers; however, their impact decreases at higher Mach numbers.

Numerical investigations were also performed by Abid et
al.~\cite{abid2022jet,abid2023influence} to study the installed jet noise
modelling and reduction using chevrons. It was found that the chevron's
penetration angle had a stronger effect on hydrodynamic pressure than the
penetration length. The near-field scattering model developed by Lyu et
al.~\cite{Lyu2017a, Lyu2018a} was implemented. The results of the implemented
model were found to be in good agreement with data for the
low-frequency noise enhancement caused by instability wave scattering for all
Mach numbers and nozzle shapes. 

Despite recent research in jet installation effects using nozzles with
chevrons, our understanding of how the chevrons change the flow and acoustics of
a jet placed nearby a wing or flap, particularly why noise reduction can be
achieved in some configurations but not others, remains inadequate. To bridge
the gap, in this paper, we conduct an experimental study to investigate the
spectral effects of chevrons on installed jet noise, its physical mechanism and
modelling using near-field spectra. Section~\ref{sec:ExpS} describes the
experimental setup, while section~\ref{sec:modelS} shows the essential
details of the scattering model developed by Lyu et
al.~\cite{Lyu2017a}. Section~\ref{sec:CompS} shows the installed jet noise
spectra and their prediction and modelling using near-field pressure statistics. Section~\ref{sec:Con} conclude the paper and lists possible future works.

\section{Experimental Methodology}
\label{sec:ExpS}

\subsection{Facility}
The acoustic measurements are conducted in the Bristol Jet Aeroacoustic Research Facility (B-JARF) at the University of Bristol. The facility, which has been validated in previous studies~\cite{kamliya2022Hydro, Screech2021, ChevronMach2021, Porous2021, Chevron2021}, allows for the generation of a clean and quiet airflow at Mach numbers above $M = 0.3$. This is accomplished through the incorporation of three specially designed in-line silencers. The first two silencers, each with a diameter of 0.3~m and a height of 1.5~m, are installed downstream of the control valve, external to the anechoic chamber. A third, larger silencer, with a diameter of 0.457~m and a height of 1.9~m, is placed inside the chamber. Perforated tubes are installed within the silencers to ensure an efficient flow passage, with the remaining space being filled with dense glass wool. The anechoic chamber itself possesses acoustic walls consisting of foam wedges and has dimensions of 7.9~m in length, 5.0~m in width, and 4.6~m in height~\cite{Mayer2019}. To minimise acoustic reflections during testing, foam is applied to the silencers, the collector, and the far-field microphone array. Total temperature and pressure measurement devices are installed within the plenum and the anechoic chamber to determine the flow conditions and the acoustic Mach number.

\subsection{Test Setup and Configuration}
\begin{figure}[!ht] 
\centering
\subfigure[SMC 000]
{
	{{ \resizebox{5cm}{!} {\fbox{\includegraphics[trim=3cm 12.5cm 19cm 2cm, clip=true]{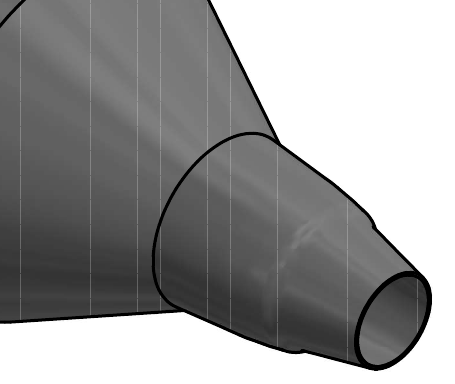}}}}}
}
\subfigure[SMC 006]
{
	{{ \resizebox{5cm}{!} {\fbox{\includegraphics[trim=3cm 12.5cm 19cm 2cm, clip=true]{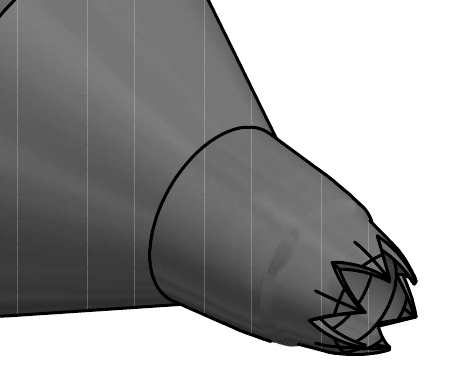}}}}}
}
\caption{Schematic of the round and chevron nozzle configurations used in the present study.}\label{fig:Nozzles}
\end{figure}

The tests are conducted using two nozzles: a round convergent nozzle (SMC000) and a chevron nozzle (SMC006), as shown in Fig.~\ref{fig:Nozzles}. The geometry of the chevron nozzle is based on the detailed analysis presented by Bridges and Brown~\cite{NASAChevron}, with the nozzles used in the present study being scaled-down versions. The SMC000 nozzle has an exit diameter of $D=16.9333$~mm. Although the tests encompass a wide range of subsonic flows, with acoustic Mach numbers from $M=0.3$ to $0.8$, the results presented herein are limited to $M=0.5$ for brevity due to similarity at other Mach numbers.

\begin{figure}[!ht] 
\centering
	{{ \resizebox{14cm}{!} {{\includegraphics[trim=1cm 2cm 1.5cm 0cm, clip=true]{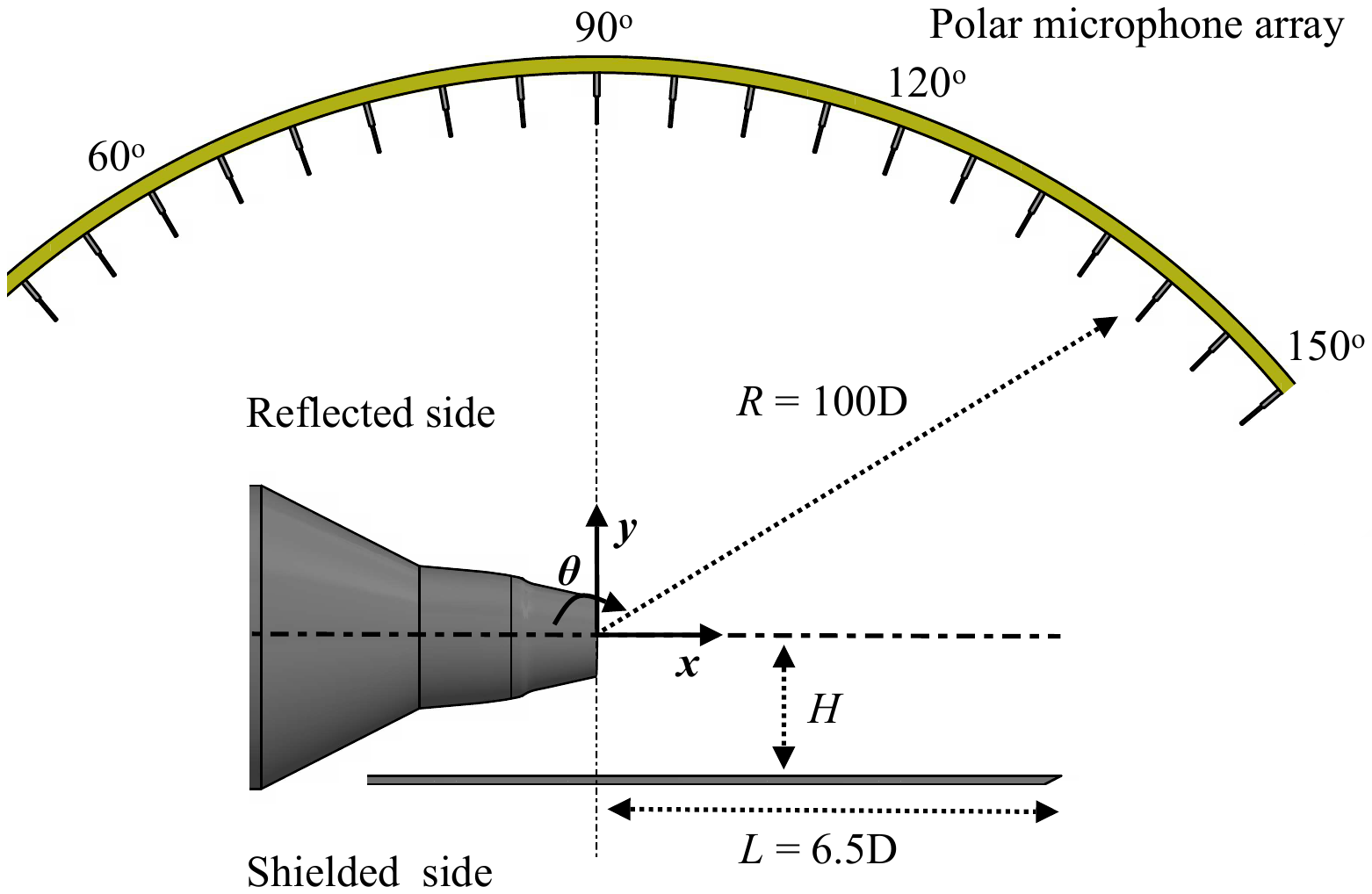}}}}}
\caption{A schematic of the experimental setup, showing the positions of the far-field microphones for the isolated and installed jet. For representational clarity, the microphones and nozzles depicted in the schematic are not drawn to scale.}\label{fig:Jet_Setup}
\end{figure}

A jet-plate configuration is used in the studies, as depicted in Fig.~\ref{fig:Jet_Setup}. The configuration consists of a 5~mm thick flat aluminium plate with a sharp trailing edge. The plate is reinforced on its underside with three 10~mm $\times$ 10~mm aluminium spars to ensure rigidity. The plate has a total length of $9.5D$ and a span of $26D$ to prevent side-edge scattering effects. To avoid scattering from the plate's leading edge, the plate extends $3D$ upstream of the nozzle exit plane. The tests are performed for an effective plate length of $L=6.5D$, as shown in Fig.~\ref{fig:Jet_Setup}. An automated traverse system is used to pos
ition the plate at various axial distances from the jet centerline to investigate the effect of plate height, i.e. $H$, as shown in Fig.~\ref{fig:Jet_Setup}. The specific non-dimensional heights, $H/D$, examined are $1.5$, $2$, $2.5$, and $3$, respectively.

\subsection{Data Acquisition and Processing}
Far-field noise measurements are acquired using an array of 21 1/4-inch G.R.A.S 40PL microphones. These microphones have a corrected flat frequency response in the range of 10~Hz to 20~kHz and a dynamic range of 150~dB. The array is positioned at a distance of 1.6~m ($R\approx100D$) from the jet exit, covering a polar angular range from $\theta=50^\circ$ to $150^\circ$, as illustrated in Fig.~\ref{fig:Jet_Setup}. For the installed configuration, the microphones are located on the reflected side. An additional far-field microphone is placed at the sideline position on the shielded side, corresponding to $\theta=270^\circ$. Data are recorded for a duration of $t=24$~s at a sampling frequency of $f_s = 2^{17}$~Hz using a National Instruments PXIe-4499 data acquisition system. The Power Spectral Density (PSD) of the acquired pressure signals is computed using a Hanning window. The data are ensemble-averaged 220 times, which yields a frequency resolution of $\Delta f = 2$~Hz. The sound pressure level (SPL) within the frequency band of $\Delta f$ is then calculated using the following equation:

\begin{equation}
\textnormal{SPL} = 10\log_{10}\left(\frac{\textnormal{PSD}(f)\Delta f}{p^2_{\textnormal{ref}}}\right),
\end{equation}

\section{Semi-analytical model of installed jet noise}
\label{sec:modelS}
The amplification of jet noise due to wing interaction is understood to be
governed by two distinct, frequency-dependent physical mechanisms. At high
frequencies, the amplification is attributed to the reflection of acoustic waves
from the wing's surface. In contrast, the significant amplification observed at
low frequencies is now understood to be due to the scattering of near-field
jet instability waves at the wing's trailing edge. 

The low-frequency amplification mechanism is of particular interest in the
present work. Accordingly, the near-field scattering model is employed to
predict this phenomenon. Note that although the model is developed in the context of round nozzles, it also applies to chevron jets as long as the plate's trailing edge is still outside the jet plume. The sound generated by this low-frequency scattering
process is considered to be largely independent of the noise produced by the
fine-scale turbulence within an isolated jet, provided the flat plate is placed
sufficiently far away from the jet so as not to touch it. Therefore, to
facilitate a direct comparison with experimental measurements of the installed
configuration, the predicted spectrum of the scattered sound is superimposed on
the spectrum of the isolated jet. Based on the instability wave scattering
mechanism at low frequencies, the power spectra density of the additional
scattering contribution in this frequency regime, $\Phi_N(\omega,
\boldsymbol{x})$, maybe be predicted by~\citep{Lyu2017a}
\begin{IEEEeqnarray}{l}
    \Phi_N(\omega, \boldsymbol{x}) = \frac{1}{\pi} \left[\frac{\omega
    x_3}{c_0 S_0^2}\right]^2\sum_{m = -N}^N \left|\frac{\Gamma(c, \mu|_{k_2
	= k\frac{x_2}{S_0}}, \mu_A)}{\mu_A}\right|^2 \Pi(\omega, m)\nonumber \\ 
  \times \left\{\sum_{k = 0}^{[\frac{|m|}{2}]} C_{|m|}^{2k} H^{- 2k + \frac{1}{2}} \gamma_c^{-|m|} \frac{d^{2k}}{dk_2^{2k}}\left[(\gamma_c^2 + k_2^2)^{\frac{1}{2}|m| - \frac{1}{4}} K_{|m| - \frac{1}{2}}\left(H\sqrt{\gamma_c^2 + k_2^2}\right) \right] \right. - \IEEEnonumber \\
  \left. \sgn(m) \sum_{k = 0}^{[\frac{|m|-1}{2}]}  C_{|m|}^{2k+1}  H^{- 2k + \frac{1}{2}}
  \gamma_c^{-|m|} \frac{d^{2k}}{d k_2^{2k}} \left[ k_2 (\gamma_c^2 +
  k_2^2)^{\frac{1}{2}|m|- \frac{3}{4}} K_{|m| - \frac{3}{2}} \left(H
  \sqrt{\gamma_c^2 + k_2^2}\right) \right] \vphantom{\sum_{k =
  0}^{[\frac{|m|-1}{2}]}}\right\}_{k_2 = \frac{kx_2}{S0}}^2, \IEEEeqnarraynumspace
  \label{equ:PhiFinal}
\end{IEEEeqnarray}
where $c$ denotes the chord of the plate used in the experiment,
($x_1, x_2, x_3$) represents the
Cartesian coordinates of the observer, the stretched distance $S_0$ is defined as
$\sqrt{x_1^2 + \beta^2(x_2^2 + x_3^2)}$, $N$ is the largest number of
azimuthal modes that need to be included, $\Pi(\omega, m)$ denotes the $m$-th
near-field pressure power spectral density, the convective radial decay rate is
given by $\gamma_c = \sqrt{(k_1 \beta^2 + k M)^2 - k^2}/\beta$, $K_i$ is
the $i$-th modified Bessel function of the second kind, $\sgn(m)$ is the sign
function, $[x]$ denotes the integer not larger than $x$, $C_{m}^{n}$ represents
the binomial coefficient and $\Gamma$ is defined by 
\begin{equation}
  \Gamma(x, \mu, \mu_A) =  \e^{\i\mu_A x} E_0(\mu x) -\sqrt{\frac{\mu}{\mu-\mu_A}}E_0\left[ (\mu-\mu_A) x \right] - \frac{1}{1 + \i}\e^{\i \mu_A x},
  \label{equ:Definitions}
\end{equation}
where 
\begin{equation}
  \begin{aligned}
      &\mu = k_1 + \sqrt{k^2-k_2^2\beta^2}/\beta^2 + kM/\beta^2, \\
      & \mu_A = k_1 + \frac{k}{\beta^2}(M - \frac{x_1}{S_0}),\\ 
      &E_0(x) = \int_{0}^{x} \frac{\e^{-\i t}}{\sqrt{2\pi t}} \ud t.
  \end{aligned}
  \label{equ:mu}
\end{equation}
Note equations~\ref{equ:PhiFinal}
also includes the effect of an ambient mean flow. For example, 
in equation~\ref{equ:mu}, $M$ denotes the Mach number of the ambient flow while $\beta
= \sqrt{1-M^2}$, $k$ denotes the acoustic wavenumber while $k_1$ and $k_2$ denote the hydrodynamic
wavenumbers of the near-field pressure in the streamwise and spanwise
directions, respectively. Since no co-flow is used in the present study, the ambient Mach number $M=0$ and $\beta=1$. 

Equation~\ref{equ:PhiFinal} is the generic form of the near-field scattering model.
However, as mentioned in \cite{Lyu2017a}, further simplifications can be achieved in practical cases. For example,
When the left- and right-rotating azimuthal modes are symmetric, i.e. $\Pi(\omega, m) = \Pi(\omega, -m)$, we can let $\Pi_s(\omega, m) =
\Pi(\omega, m) + \Pi(\omega, -m)$ for $m \ne 0$ to represent the single-sided spectra density and write the far-field sound spectral density in the mid-span plane
($x_2 = 0$) as
\begin{equation}
    \Phi_N(\omega, \boldsymbol{x}) \approx \left[\frac{\omega x_3}{c_0
    S_0^2}\right]^2 \left\{\left|\frac{\Gamma(c, \mu, \mu_A)}{\mu_A}\right|^2
    \frac{\e^{-2H\gamma_c}}{2\gamma_c^2}\right.
    \left.\left(\frac{\Pi_0(\omega, 0)}{K_0^2(\gamma_c r_0)}+ \frac{\Pi_0(\omega, 1)}{K_1^2(\gamma_c r_0)}\right)\right\}_{U_c = \overline{U}_c(\omega),k_2=0},  \IEEEeqnarraynumspace
    \label{equ:PhiMidSpanSemifinal}
\end{equation}
where $\Pi_0(\omega, m)$ is the $m$-th single-sided spectrum of the incident
near-field evanescent instability waves measured at a radial distance $r = r_0$,
and $U_c$ is the convection velocity of these waves. In practice, the radial
location $r_0$ is chosen to be sufficiently small such that the measurement is
located within the hydrodynamic near-field of the jet, where acoustic
fluctuations can be considered negligible. The model requires the spectra for
the axisymmetric ($m=0$) and first helical ($m=1$) modes as inputs. For the
present study, these spectral data are sourced from the author's previous Large
Eddy Simulation (LES) for both round and chevron jets~\cite{kamliya2021effects}. As the spectra vary with axial
position, the data are extracted at a location corresponding to the intended
axial position of the plate's trailing edge. Furthermore, a frequency-dependent
convection velocity, $\overline{U}_c(\omega)$, also obtained from earlier LES
studies~\cite{Lyu2017a}, is used. Note that this noise prediction model can also be extended to predict the scattered jet noise when the flat plate has a swept trailing edge; details are not used here, but can be found from \cite{Lyu2018h,Lyu2019a}.

Equation~\ref{equ:PhiFinal} is expected to predict the far-field installed jet noise spectra for both round and chevron jets, provided the plate is situated within the hydrodynamic near-field of the jet, a condition that is generally satisfied in the present study. It is noted, however, that the use of chevrons results in a faster jet spread. Consequently, the reference position $r_0$ for the chevron nozzle case is chosen to be slightly larger than that for the round nozzle. As previously stated, the predicted near-field scattering contribution, $\Phi_N$, is superposed on the measured isolated jet noise to facilitate a direct comparison between the model and experiments. Given its dominance at low frequencies, the scattering contribution is expected to reproduce the low-frequency amplification observed in the experimental data. Conversely, as the scattered contribution decays rapidly with increasing frequency, the total predicted spectrum is expected to collapse onto the isolated jet spectrum at high frequencies. The model does not account for the high-frequency noise increase (typically around 3~dB) caused by acoustic reflection from the plate surface, as this mechanism is well-understood and is not the focus of the current predictive effort.

\section{Results and Discussion} \label{sec:CompS}
\subsection{Far-field noise validation}

\begin{figure}[!ht] 
\centering
\subfigure[SMC000 Isolated]
{
    {{ \resizebox{8cm}{!} {{\includegraphics[trim=0cm 0cm 0cm 0cm,
    clip=true]{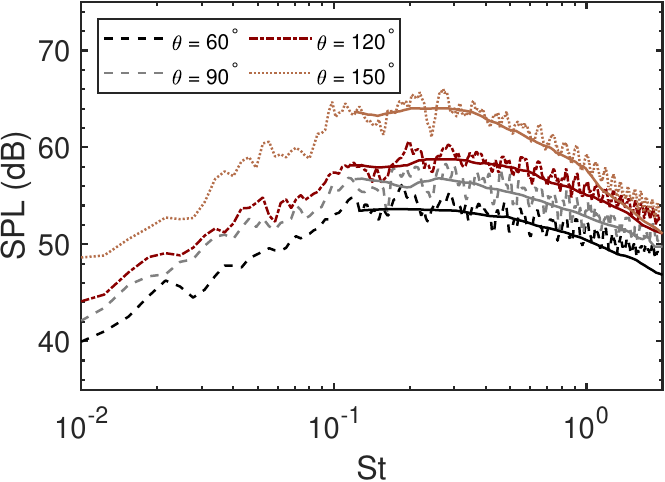}}}}}
}
\caption{Comparison of SPL for isolated round convergent nozzle (dashed line) with experimental data available in the literature Brown and Bridges \cite{brown2006small}  (solid lines) for select polar angles ($\theta = 60^\circ, 90^\circ, 120^\circ$ and $150^\circ$) obtained at acoustic Mach number $M = 0.5$.}\label{fig:Vali}
\end{figure}

There often exists much uncertainty in the jet noise spectra measurements at low
Mach numbers, as previously discussed in the
literature~\cite{viswanathan2003jet,viswanathan2006instrumentation}. Therefore,
to validate the accuracy of the measurements in this study, the far-field
results are first compared with those from previous studies with similar flow
conditions. Specifically, the results from Brown and
Bridges~\cite{brown2006small} and Bastos \textit{et
al.}~\cite{bastos2018development} are used to compare with the far-field measurements
taken at various inlet angles ($\theta$) for a round nozzle (SMC000). This
present nozzle has an exit diameter of $D=16.933$~mm, and the Mach number is $M=0.5$,
resulting in a Reynolds number of $Re=196,000$. Upon comparison, it can be seen 
that the results from the current experimental study align well with those
from the literature, after accounting for distance and diameter factors, as
shown in Fig.~\ref{fig:Vali}. This consistency demonstrates that the current
experimental setup is capable of obtaining accurate spectral measurements for
low Mach number jets.

\subsection{Far-field spectra}

\begin{figure}[!ht] 
\centering
	{{ \resizebox{\figwidth}{!}{{\includegraphics[trim=0cm 21.4cm 9.5cm 0cm, clip=true]{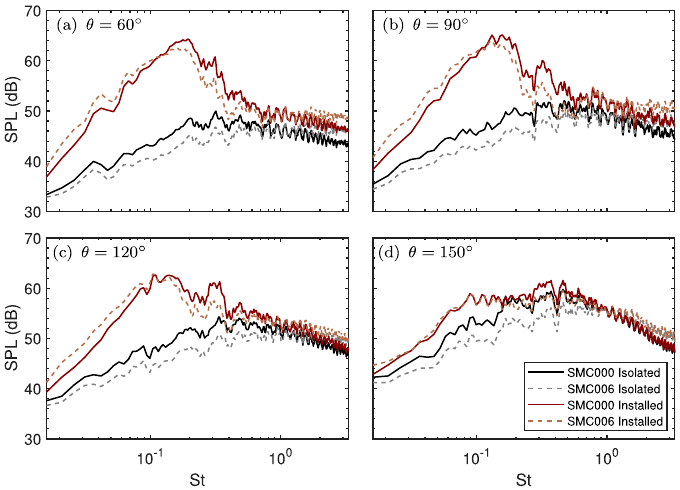}}}}}
\caption{Sound pressure level comparison for round and chevron nozzles measured
at various polar angles $\theta=60^\circ,90^\circ,120^\circ$ and $150^\circ$ for
the isolated and installed configurations at acoustic Mach number of
$M=0.5$.}\label{fig:FirstSPL_05}
\end{figure}

The sound pressure level (SPL) for both isolated and installed configurations of
chevron and round jets at directivity angles $\theta = 60^{\circ}$,
$90^{\circ}$, $120^{\circ}$, and $150^{\circ}$ are shown in
Fig.~\ref{fig:FirstSPL_05}. These SPL results are plotted against the Strouhal
number based on the jet diameter $D$ ($St = fD/U_j$), providing a comprehensive overview of the acoustic characteristics of the jets under different configurations.

First, the isolated jet noise spectra are shown by the black lines in Figs.~
\ref{fig:FirstSPL_05}(a-d). The results for the chevron (SMC006) isolated
configuration demonstrate a significant reduction in noise levels compared to
round nozzles. The reduction is approximately 3-5 dB and occurs primarily at low
to mid frequencies across all presented directivity angles. This substantial
noise reduction highlights the effectiveness of the chevron design in mitigating
acoustic emissions in these frequency ranges. However, it should also be noted
that chevron nozzles show a notable increase in high-frequency noise for $St >
1.5$. This increase is attributed to the generation of smaller-scale turbulent
structures caused by the streamwise vortices introduced by the chevron design.
The noise reduction for the SMC006 chevron isolated configuration at low to mid frequencies is expected and is due to the enhanced mixing of the jet with the
ambient air, weakening large-scale turbulent structures, as previously observed
by Bridges and Brown~\cite{NASAChevron}.

Second, the installed noise spectra are shown by the red lines in Fig.~\ref{fig:FirstSPL_05}(a-d). Data in solid lines represent noise spectra for round jets. As expected, the results demonstrate a distinctive low-frequency spectral hump in the $St=0.01-0.5$ range. This low-frequency amplification is most pronounced at upstream and side angles and gradually diminishes at downstream angles closer to the jet axis at $150^\circ$. This amplification at low frequencies is attributed to the hydrodynamic interaction between the jet flow and the trailing edge of the installation plate. These interactions enhance the low-frequency noise due to the scattering of near-field hydrodynamic waves at the solid trailing edge, as observed in previous studies~\cite{Lyu2017a}. As the observation angle moves downstream, the influence of these interactions decreases, resulting in a reduction in the low-frequency amplification. This behaviour is consistent with the dipolar directivity pattern, where the maximum energy is radiated in the direction normal to the scattering surface~\cite{Head1976,Cavalieri2014}. 

Interestingly, the installed configuration of the chevron nozzles (SMC006) shows
a considerable reduction in SPL of approximately 2-3 dB in the $St=0.15-0.5$
range and an increase in SPL in the $St=0.01-0.15$ range except at the
downstream angle of $150^\circ$. The noise increase between the Strouhal number
of $0.01$ and $0.15$ is most prominent at upstream and sideline angles. The
reduction in SPL at $St=0.15-0.5$ can be attributed to the chevron nozzles'
ability to enhance mixing and break down large-scale turbulent structures,
resulting in less acoustic energy being radiated at these frequencies.
Conversely, the increase in SPL at $St=0.01-0.15$ is likely due to the faster
spreading of the jet due to enhanced mixing, resulting in a shorter distance
between the jet plume and the flat plate's trailing edge. The near-field
pressure fluctuations near the flat plates' trailing edges are therefore higher for the
chevron case, leading to louder noise compared to round jets. The fact that such a
noise increase becomes negligible at $150^\circ$ is likely to be caused by the
significant refraction of the sound by the jet mean flow at downstream angles.
Such a refraction does not present at $90^\circ$, explaining the difference from
those at $150^\circ$. Overall, the installed configuration of the SMC006 nozzle
exhibits higher noise levels compared to the SMC000 nozzle at nearly all 
observer angles when the Strouhal number is between $0.01$ and $0.15$.
Conversely, in isolated configurations, the SMC006 nozzle demonstrates lower
noise levels than the SMC000 nozzle.

\begin{figure}[!ht] 
\centering
	{{ \resizebox{14cm}{!}{{\includegraphics[trim=0cm 13.7cm 9.4cm 0cm, clip=true]{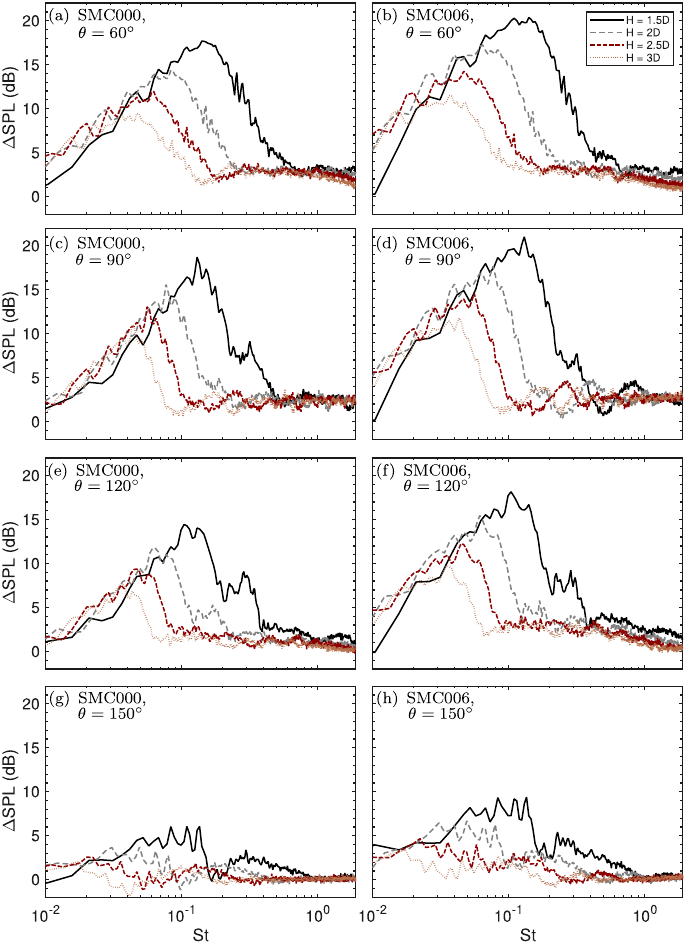}}}}}
    \caption{Sound spectral difference $\Delta$SPL for various plate heights and  observer angles.}\label{fig:SPL_Height_90}
\end{figure}

To perform a comprehensive analysis of the impact of plate height $H$ on the
installation noise, Fig.~\ref{fig:SPL_Height_90} presents the $\Delta$SPL
spectra for both round (SMC000) and chevron (SMC006) nozzles at plate heights
ranging from $H/D=1.5$ to $3$. The spectra are shown at $M=0.5$ for an observer
angle of $\theta=60^\circ$, $90^\circ$, $120^\circ$ and $150^\circ$ in
Figs.~\ref{fig:SPL_Height_90}(a-b), (c-d), (e-f) and (g-h), respectively. The
$\Delta$SPL is calculated relative to the corresponding isolated jet. All
spectra clearly exhibit a low-frequency broadband hump caused by installation
effects, which diminishes in magnitude as the plate height increases. As the
plate height increases, the spectral hump shifts to lower frequencies; for
instance, the hump covers a wide region between $St=0.01-0.5$ for $H = 1.5D$,
and narrows to $St=0.01-0.3$ for $H = 2D$, $St=0.01-0.2$ for $H = 2.5D$, and
further reduces to $St=0.01-0.15$ for $H = 3D$. This blue shift is due to the
variation of the near-field pressure fluctuations. It is known that
high-frequency pressure fluctuations decay faster in the radial direction
compared to those at low frequencies. Hence, as $H$ increases, the pressure
fluctuations are not only weaker but also increasingly dominated by
low-frequency components. Such a change is reflected clearly in the installed
noise spectra. 

It is interesting to note that a secondary peak emerges within the broadband
hump for both tested nozzles at the closest plate height of $H = 1.5D$ at
$\theta=90^\circ$ and $120^\circ$. A reasonable hypothesis for its appearance is
due to the interference between the direct sound generated by large coherent
structures and that generated due to the scattering of the near-field pressure
fluctuations. At such a short distance, the plate is virtually in direct contact
with the plate's trailing edges. The pressure fluctuations due to the large
coherent flow structures are likely directly scattered into sound. In the meantime, they are emitting sound directly. The two sounds are
likely to interfere in the far field, leading to the appearance of spectral
oscillation and hence secondary peaks. As $H$ increases, the near-field pressure
fluctuations are increasingly less coherent with the far-field sound generated
directly by the large coherence structures, hence explaining the disappearance of
such a secondary peak. In addition, at upstream observer angles, the noise
contribution of the large coherent structures is known to be weak, therefore
explaining why no spectral peaks are found at $\theta=60^\circ$. This hypothesis
would be further corroborated by analyzing the coherence between far-field sounds
in section~\ref{subsec:coherence}. Note that this observation aligns with previous
findings by Jawahar et al. \cite{Porous2023}. 

The impact of plate height on the
spectra extends beyond $H/D=2$, consistent with the extended linear hydrodynamic
field demonstrated by Suzuki and Colonius \cite{suzuki2006} for $St=0.25-0.50$
and azimuthal modes ($m=0$, $\pm1$, and $\pm2$). Comparing
Fig.~\ref{fig:SPL_Height_90}(a,c,e,g) with (b,d,f,h) shows that the SMC006 configuration shows a considerable increase in noise compared to the SMC000
configuration at all tested plate heights and presented directivity angles. In
addition, the spectral hump appears wide than that for SMC000. In other words,
the installation effects appear stronger in the chevron case. Such a difference
from that for the round nozzle is likely due to the faster spread of the jet
plume due to the enhanced mixing for the chevron nozzle.

\subsection{Overall sound pressure level trends}

\begin{figure}[!ht] 
\centering
	{{ \resizebox{16cm}{!}{{\includegraphics[trim=0cm 20.4cm 8.1cm 0cm, clip=true]{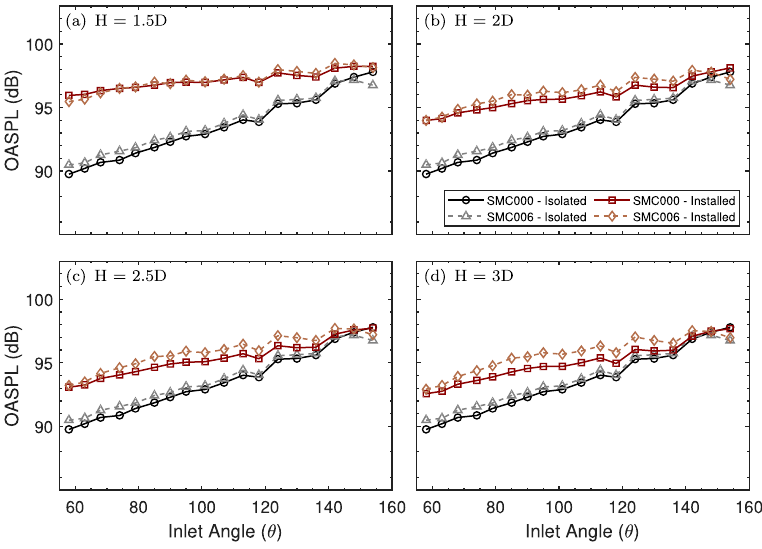}}}}}
\caption{OASPL ($St = 0.01-3$) at different directivity angles for both isolated and installed configurations across various plate heights .}\label{fig:OASPL_Height}
\end{figure}

The Overall Sound Pressure Level (OASPL) was computed over a frequency range
corresponding to $St=0.01-3$ using a far-field microphone array positioned at
various polar angles. This analysis was performed for both SMC000 and SMC006
under both isolated and installed configurations at a Mach number of $M=0.5$.
Fig.~\ref{fig:OASPL_Height} presents the OASPL results at various plate heights
for both configurations. Each plot compares the round and chevron nozzles. The
overall trend for the isolated jet indicates a gradual increase in OASPL as the
observer moves from upstream to downstream; this is expected because isolated
jet noise is more pronounced at low observer angles. The results for the
installed configuration show a substantial increase of 5~dB in the OASPL in the
upstream angles ($\theta<90^\circ$). This increase is attributed to the increase
in low-frequency amplification due to the hydrodynamic interaction with the flat
plate. The OASPL distribution along the inlet angle $\theta$ is therefore much
flatter compared to that of the isolated jet. In addition, as the flat height
$H$ increases, the increase in OASPL at upstream and sideline angles is
substantially reduced. This is expected due to the much weaker noise
amplification for large $H$ values at low frequencies. For the isolated condition, the
results show a mild increase in OASPL for the chevron (SMC006) configuration
compared to the round nozzle. This could be attributed to the increase in the
high-frequency noise that arises for the chevron configuration, which
contributes more energy to the OASPL than that at low frequencies. For the installed configuration, the
chevron nozzle shows a more noticeable increase in the OASPL compared to the
round jet except at $H=1.5D$. Again, this is likely due to the faster spread of
the chevron jet leading to stronger near-field pressure fluctuations. At
$H=1.5D$, the jet plume might be in direct contact with the plate, causing
deviation from this tendency.


\begin{figure}[!ht] 
\centering
	{{ \resizebox{16cm}{!} {{\includegraphics[trim=0cm 25.7cm 10.5cm 0cm, clip=true]{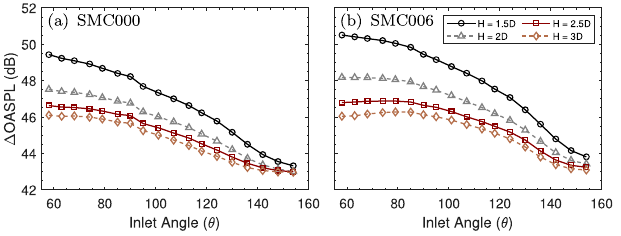}}}}}
\caption{Delta OASPL ($St = 0.01 - 3$)  at different directivity angles for both round and chevron configurations across various plate heights at $M=0.5$.}\label{fig:OASPL_D}
\end{figure}

Fig.~\ref{fig:OASPL_D} displays the difference in OASPL ($\Delta$OASPL) between
the installed and isolated configurations at $M=0.5$. The OASPL used to
calculate the $\Delta$ values was determined over a frequency range
corresponding to $St=0.01-3$. The results are categorised by the plate height,
with the data for round nozzles shown on the left and the chevron nozzles on the
right. This comparison highlights the impact of installation effects across
different nozzle types and plate heights. From Fig.~\ref{fig:OASPL_D}, it is
clear that the chevron nozzle exhibits higher noise levels compared to the round
nozzle. The highest levels of $\Delta$OASPL are observed at low inlet angles
($\theta < 90^\circ$), with the difference substantially reduced as $\theta$
approaches $150^\circ$. This trend is expected as discussed in
Fig.~\ref{fig:OASPL_Height}. Increasing the plate height further reduces
$\Delta$OASPL. Additionally, the change in noise levels as $H$ varies is more
pronounced for the chevron nozzle configuration compared to the round nozzle.
For instance, at upstream angles, the SMC006 chevron nozzle shows a 5~dB
variation between the first two plate heights ($1.5D-2.5D$), whereas the SMC000
round nozzle exhibits a 4~dB difference. Although chevron nozzles are typically
known for noise reduction in isolated settings, in the installed jet
configuration, the enhanced mixing leads to stronger hydrodynamic interactions
with nearby surfaces, causing more pronounced installation effects.

\subsection{Coherence}
\label{subsec:coherence}

Following previous studies~\cite{Mead1998,Cavalieri2014, Lyu2017a}, which suggest that the low-frequency noise amplification is due to the scattering of the near-field hydrodynamic field by the plate's trailing edge, this study aims to examine this scattering process more closely by computing the far-field sound coherence for both round and chevron nozzles. To investigate this, pressure signals are analysed from microphones placed on opposite sides of the trailing edge. Specifically,  one microphone at $\theta=270^\circ$ on the shielded side is used, and coherence is studied between this microphone with various microphones on the reflected side. The coherence function between these microphone pairs is then calculated to assess the importance of this scattering process. The coherence was calculated using the following equation 
\begin{equation}
\centering
\gamma_{p_ip_j}^2(f)=\dfrac{\mid \Phi_{p_ip_j}(f)\mid^2}{\Phi_{p_ip_i}(f)\Phi_{p_jp_j}(f)}~~\mbox{for}~~p_i \text{ (}  {60^\circ, 90^\circ, 120^\circ, 150^\circ})~\mbox{and}~p_j \text{ (} {270^\circ)},
\label{Eq:FFCoh}
\end{equation}
where $\Phi_{p_ip_j}(f)$, $\Phi_{p_ip_i}(f)$ and $\Phi_{p_jp_j}(f)$ represent the cross-PSD between $p_i$ and $p_j$, the PSDs of $p_i$ and $p_j$, respectively. 

\begin{figure}[!ht] 
\centering
	{{ \resizebox{16cm}{!} {{\includegraphics[trim=0cm 21.5cm 9.8cm 0cm, clip=true]{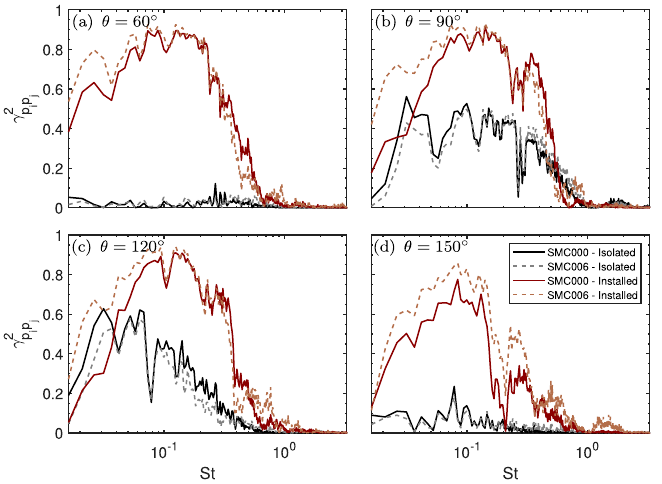}}}}}
\caption{Coherence between the $\theta=270^\circ$ and various far-field microphones on the reflected side for $M = 0.5$ at $H = 1.5D$.}\label{Fig:Coh05}
\end{figure}

The normalized coherence results for the different configurations at $\theta = 60^\circ$, $90^\circ$, $120^\circ$, and $150^\circ$ are presented in Fig.~\ref{Fig:Coh05}. These results reveal several important trends that are consistent with previous studies on jet noise and jet installation effects.

First, it is evident that for both nozzle types, the installed configurations
exhibit significantly higher coherence levels across the Strouhal number range of 0.01 and 0.5 compared to the isolated cases. Specifically, for the round nozzle (SMC000), coherence levels in the installed configuration reach values as high as $\gamma_{p_ip_j}^2 > 0.8$ for $M=0.5$ across multiple polar angles,
particularly in the low-frequency range. In contrast, the isolated configuration displays much lower coherence values, typically ranging from $\gamma_{p_ip_j}^2 = 0.2-0.4$, depending on the angle and frequency. This non-zero baseline coherence in the isolated jet is likely due to the sound radiated by large-scale coherent turbulent structures within the jets. These structures are known to radiate efficiently at low frequencies, resulting in relatively large azimuthal coherence. However, this coherence is significantly lower than that in the installed configuration, where the trailing edge acts as a single, compact scattering point, enforcing a much stronger phase relationship between the shielded and reflected sound fields. 


For the chevron nozzle (SMC006), a similar trend is observed, although the coherence levels are generally higher than those of the round nozzle, indicating that chevrons, despite their noise-reduction capability in isolated conditions, enhance scattering when installed, leading to increased coherence. This is particularly notable at the sideline angles (around $90^\circ$), where the interaction noise between the jet and the surface is most pronounced due to the dipolar directivity. The increased coherence in the chevron nozzle supports the hypothesis that the enhanced mixing and turbulence introduced by the chevrons lead to stronger interactions with nearby surfaces, thus amplifying the scattered noise.

As the polar angle increases to $\theta = 120^\circ$ and $150^\circ$, the
coherence levels in the installed configuration decrease, reflecting the reduced contribution of the scattering sound to the total noise at these angles. Again,
this is expected due to the dipolar directivity of the scattering sound. Note
that the coherence for the isolated configuration also decreases at large
observer angle of $\theta=150^\circ$. Furthermore, the presence of a secondary spectral hump in the
coherence plots, particularly at $\theta = 90^\circ$, indicates a dipole-type
noise source; this suggests that the secondary hump appears in
Fig.~\ref{fig:SPL_Height_90} is also due to a similar scattering mechanism as
that in lower frequencies, supporting the hypothesis that it is formed by the
constructive interference between the direct sound and the sound arising from the scattering.

In summary, the coherence analysis across different angles and configurations demonstrates the significant impact of installation effects on jet noise, particularly in the low-frequency range. The chevron nozzles, while effective in isolated noise reduction, exhibit higher sound coherence under installed configurations, suggesting an increased scattering effect. This insight is crucial for understanding and mitigating installation noise in practical aerospace applications.

\begin{figure}[!ht] 
\centering
	{{ \resizebox{16cm}{!} {{\includegraphics[trim=0cm 25.3cm 9.7cm 0cm, clip=true]{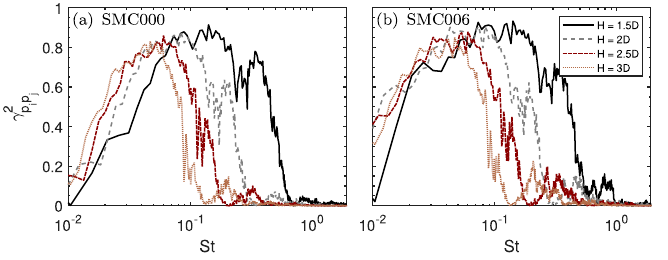}}}}}
\caption{Coherence between the $\theta=270^\circ$ and $\theta=90^\circ$  for various plate heights at $M = 0.5$.}\label{Fig:Coh_h}
\end{figure}

To fully characterise the scattering mechanism at other plate positions, coherence
analysis is also carried out between two far-field microphones at polar angles
$\theta=90^\circ$ and $270^\circ$ for the plate height $H$ of $1.5D$, $2D$,
$2.5D$ and $3D$, respectively. The results for the isolated and installed configurations are presented in Fig.~\ref{Fig:Coh_h}. One clear trend is that the coherence
tends to decrease slightly as $H$ increases. This is expected since the far-field sound is increasingly less dominated by the scattering. In addition, the
bandwidth of the large coherence starts to decrease and the peak coherence
frequency shifts to lower frequencies; this is consistent with the well-known blue shift of the installation effects, as also demonstrated in Fig.~\ref{fig:SPL_Height_90}. Note also that the secondary humps are only evident at $H=1.5D$ and $2D$ because the
interference is weaker at large jet-plate separation lengths.

\section{Prediction and Comparison with Experiments}
In this section, the far-field noise spectra obtained using the prediction model shown in equation~\ref{equ:PhiMidSpanSemifinal} are compared with experimental results. Comparisons for both round and chevron nozzles are shown. As introduced in section~\ref{sec:modelS}, only the 0-th and 1-st order modes of the near-field instability waves are included.  

\begin{figure}[!ht] 
\centering
	{{ \resizebox{16cm}{!}{{\includegraphics[trim=0cm 17.5cm 9.4cm 0cm, clip=true]{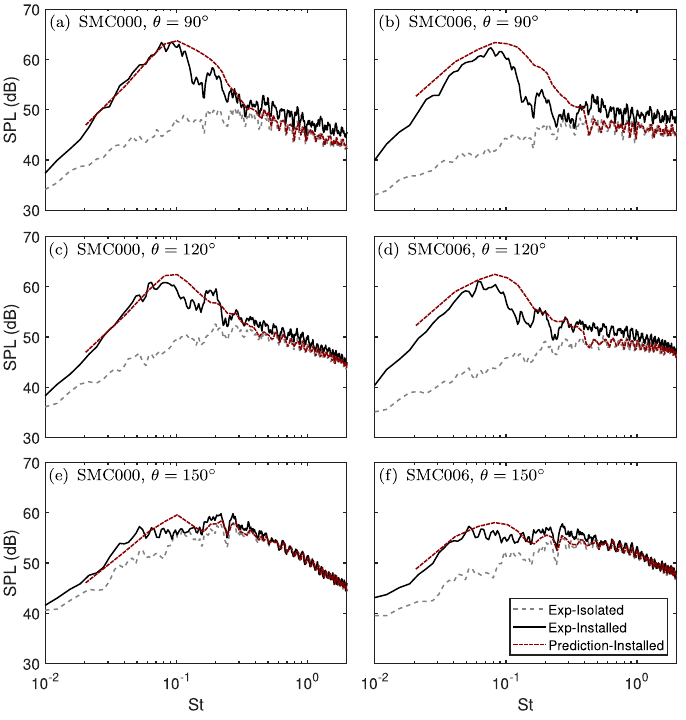}}}}}
\caption{Comparison between experimental and installed jet noise prediction for round and chevron jets when $H/D=1.5$.}
\label{fig:modelH1_5D}
\end{figure}

Fig.~\ref{fig:modelH1_5D} shows the comparison when the plate is at $H/D=1.5$. At this close distance, significant noise amplification is expected, which is evident in Fig.~\ref{fig:modelH1_5D}. For example, Fig.~\ref{fig:modelH1_5D}(a) shows a noise increase of up to 20~dB for round nozzles for an observer at $90^\circ$. Such strong noise amplification is well predicted by the model. Note again that the secondary hump appearing at around $St=0.2$ is possibly due to interference between the direct and scattered sounds. The model does not include such an interference and therefore cannot predict such a hump; however, the amplitude follows closely with the experimental value. When the observer moves to $120^\circ$, a weaker noise amplification is observed. This results from the dipolar directivity of the installed jet noise at low frequencies. The model can capture this trend and agrees with the experimental spectrum in both spectral shape and amplitude. As the observer moves to $150^\circ$, the noise increase is limited to around 5 dB. This is well predicted by the model.

When the chevron nozzle is used, the near-field noise amplification appears
stronger, e.g. up to 25~dB as shown in Fig.~\ref{fig:modelH1_5D}(b). This is
partly due to the stronger installation effects due to enhanced mixing (hence stronger near-field pressure fluctuations at a fixed radial position), partly because noise suppression occurs for isolated jet noise when chevrons are used. Comparing Figs.~\ref{fig:modelH1_5D}(b) and (a) also reveals the reduction in isolated jet noise at low frequencies. Fig.~\ref{fig:modelH1_5D}(b) shows that the prediction model continues to work for chevron nozzles. A small overprediction by the model is observed, which is likely because the chevron enhances the jet
mixing, resulting in a direct contact between the jet plume and the flat plate.
Hence for a more accurate predition may be achieved by using the actual pressure
fluctuation spectrum at the position within the jet plume where the flat plate's
trailing edge is. Nevertheless, Fig.~\ref{fig:modelH1_5D}(b) shows that the
model described in section~\ref{sec:modelS} is still applicable for installed
chevron jets. The trend that increasing the observer angle weakens the
low-frequency noise increase, and the good agreement between predictions and
experiments remains similar to those for round jets. We therefore omit a
repetitive description.

\begin{figure}[!ht] 
\centering
	{{ \resizebox{16cm}{!}{{\includegraphics[trim=0cm 17.5cm 9.4cm 0cm, clip=true]{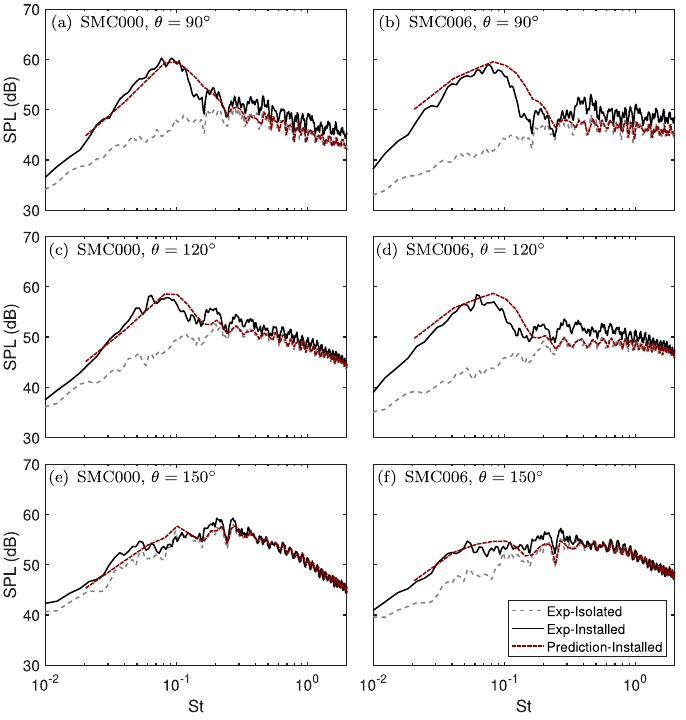}}}}}
\caption{Comparison between experimental and installed jet noise prediction for round and chevron jets when $H/D=2D$.}\label{SPL_05}
\label{fig:modelH2D}
\end{figure}
Moving the plate slightly far away to $H=2D$ results in a weaker noise amplification, as shown in Fig.~\ref{fig:modelH2D}. This is because the plate's trailing edge is located in the linear hydrodynamic regime of the jet, where the near-field pressure fluctuations decay exponentially~\cite{Jordan2013}. For example, the maximum noise increase shown in Fig.~\ref{fig:modelH2D}(a) reduces to 15 dB, compared to 20 dB shown in Fig.~\ref{fig:modelH1_5D}(a). The model can capture this behaviour very well, and shows good agreement with the experimental spectrum. As the observer angle increases to $120^\circ$ and $150^\circ$, 
the dipolar directivity leads to increasingly weak noise amplification. The installed jet noise model predicts this behaviour well. The larger noise amplification that occurs in the chevron nozzle case appears largely similar to Fig.~\ref{fig:modelH1_5D}. Figs.~\ref{fig:modelH2D}(b,d,f) show that the model can predict the installed spectra for the chevron nozzle accurately, further demonstrating the applicability of the near-field scattering model for chevron nozzles. 

\begin{figure}[!ht] 
\centering
	{{ \resizebox{16cm}{!}{{\includegraphics[trim=0cm 17.5cm 9.4cm 0cm, clip=true]{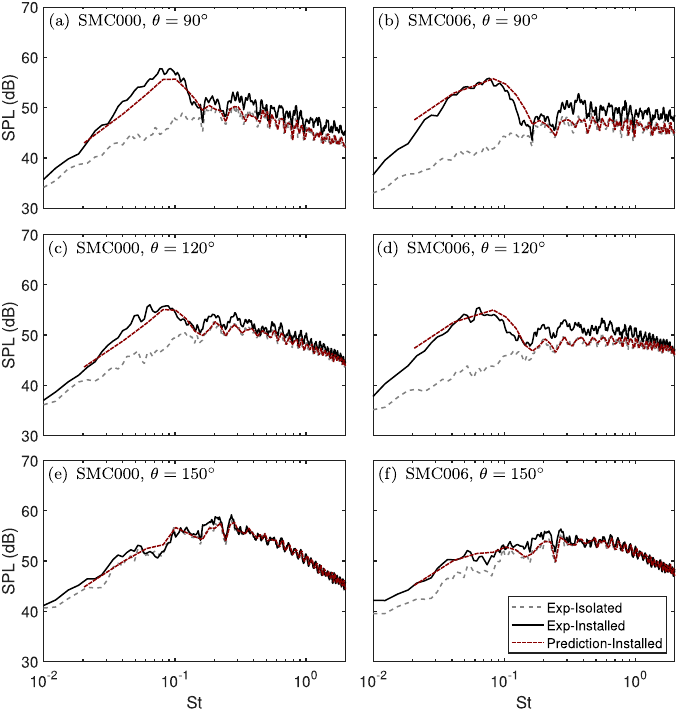}}}}}
\caption{Comparison between experimental and installed jet noise prediction for round and chevron jets when $H/D=2.5$.}
\label{fig:modelH2_5D}
\end{figure}
When the plate is moved to $H=2.5D$, the same trend continues. As shown in
Fig.~\ref{fig:modelH2_5D}(a), the maximum noise enhancement is around 12 dB
and the low-frequency peak appears narrower. The model predicts a low-frequency
peak of similar shape with a maximum noise increase of 10 dB. When the observer
is around $120^\circ$, this noise amplification decreases to 10 dB. Furthermore,
when the observer is located at $150^\circ$, virtually no noise amplification is
observed, agreeing well with the predictions. This is because at this location
and observer angle, the noise due to the near-field scattering is lower than
the contributions from quadrupole sources. Hence, the total noise spectrum
looks virtually identical to the isolated spectrum.  This is more evident in the
chevron case shown in Fig.~\ref{fig:modelH2_5D}(f), where noise increase is
still distinguishable because the isolated noise energy is lower and the
near-field pressure fluctuations are stronger than those for the round nozzle.
Similar to Figs.~\ref{fig:modelH1_5D} and \ref{fig:modelH2D}, the model is
still capable of yielding a good prediction for chevron nozzles when the
observer is at $90^\circ$ and $120^\circ$, as shown in
Fig.~\ref{fig:modelH2_5D}(b) and (d), respectively.

\begin{figure}[!ht] 
\centering
	{{ \resizebox{16cm}{!}{{\includegraphics[trim=0cm 17.5cm 9.4cm 0cm, clip=true]{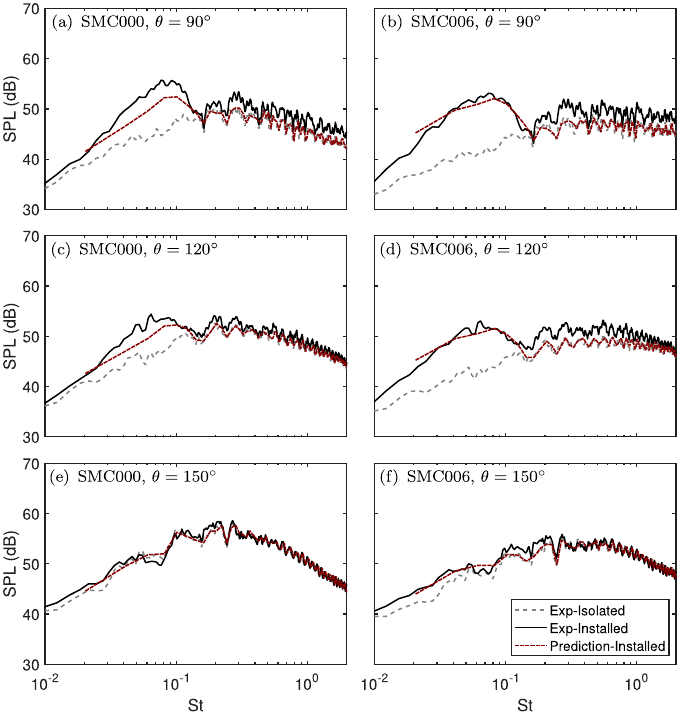}}}}}
\caption{Comparison between experimental and installed jet noise prediction for round and chevron jets when $H/D=3D$.}
\label{fig:modelH3D}
\end{figure}

Fig.~\ref{fig:modelH3D} shows the comparison at the largest vertical separation distance of $H=3D$. This is farthest from the centerline. The near-field pressure fluctuations are therefore of low amplitudes. The maximum noise increase is around 10 dB at $90^\circ$ and reduces to 3 dB and 0 dB for $\theta=120^\circ$ and $150^\circ$, respectively. The model predicts a slightly lower amplitude compared to experiments, possibly due to an inaccurate frequency-dependent convection velocity used in the model. The predictions for the chevron nozzles, however, appear good across all observer angles. At this large separation distance, no noise increase occurs for $\theta=150^\circ$ even for the chevron case, as can be seen in Fig.~\ref{fig:modelH3D}(f). 

In summary, Figs.~\ref{fig:modelH1_5D} to \ref{fig:modelH3D} show that although the chevron nozzle enhances jet mixing and results in stronger jet installation effects, and the scattering-based prediction model works well in predicting the low-frequency amplifications. Small deviations might occur at the shortest jet-plate distance of $H=1.5D$, possibly due to a direct contact between the jet plume and the flat plate's trailing edge. 

\section{Conclusion and future work}
\label{sec:Con}

This paper studies the impact of chevrons on installed jet noise by
measuring the far-field sound at a jet Mach number of $0.5$. Both round (NASA
SMC000) and chevron (SMC006) nozzles are used; The SMC000 and SMC006 nozzles of
a diameter $D=16.93$~mm are placed near a flat plate. The jet is
separated from the flat plate by a horizontal distance $L=6.5D$ between the
plate's trailing edge and the nozzle exit. The vertical separation distance $H$
varies between $1.5D$, $2D$, $2.5D$ and $3D$. 

Far-field sound is measured on a microphone arc covering an observer angle range
from $60^\circ$ to $150^\circ$ to the downstream jet axis. Measurements are
conducted on both the reflected and shielded sides. The measured sound spectra
are compared to the near-field scattering model developed by Lyu et
al.~\cite{Lyu2017a}, together with the near-field spectra inputs obtained from
LES. To further understand the physical mechanism of the noise amplification at
low frequency for both round and chevron nozzles, a coherence analysis is
conducted between one microphone placed on the reflected side ($60^\circ$,
$90^\circ$, $120^\circ$ and $150^\circ$ to the downstream jet axis) and the
other on the shielded side ($270^\circ$ to the downstream jet axis).

Examining the far-field sound spectra shows that chevrons reduce isolated jet
noise at low frequencies, but also increase noise at high frequencies,
consistent with previous findings. A comparison between the isolated and
installed spectra shows that jet installation results in a strong noise
amplification at low frequencies due to the scattering of the near-field
pressure fluctuations. In addition, a mild noise increase on the reflected side
is observed at high frequencies due to the noise reflection off the plate's
surface. Noise reduction can be achieved on the shielded side. 

The low-frequency amplification is most significant at $H=1.5D$, since the
near-field pressure fluctuations are strongest at this position.  The installed
jet noise has a dipolar directivity at low frequencies, hence it is most
significant at sideline angles (close to $\theta=90^\circ$). The use of
chevrons leads to even stronger noise amplification; this is likely due to the
enhanced jet mixing resulting in a faster spreading jet plume and the
corresponding stronger near-field pressure fluctuations at a fixed radial distance. 

It is interesting to observe that a secondary spectral peak appears within the
primary low-frequency noise increase hump. It is hypothesised that this secondary peak is due to the interference between the sound generated directly
by the large coherent structures and by their scattering. The comparison
between the predicted and measured far-field spectra shows that the scattering model can reproduce the low-frequency noise amplification well at various
observer angles for both round and chevron jets, suggesting the validity of the
instability-wave scattering mechanism for both round and chevron jets.

\begin{acknowledgments}
The first and third authors (H.J and M.A.) would like to acknowledge the Engineering
and Physical Sciences Research Council (EPSRC) for supporting this research
(Grant No. EP/S000917/1). The second author (B.L) would like to acknowledge the
funding supported by the National Natural Science Foundation of China (Grant No
12472263) and the Beijing Natural Science Foundation (Grant No. L253027). 
\end{acknowledgments}

\bibliography{Ref_clean_HKJ}

\begin{thebibliography}{46}%
\makeatletter
\providecommand \@ifxundefined [1]{%
 \@ifx{#1\undefined}
}%
\providecommand \@ifnum [1]{%
 \ifnum #1\expandafter \@firstoftwo
 \else \expandafter \@secondoftwo
 \fi
}%
\providecommand \@ifx [1]{%
 \ifx #1\expandafter \@firstoftwo
 \else \expandafter \@secondoftwo
 \fi
}%
\providecommand \natexlab [1]{#1}%
\providecommand \enquote  [1]{``#1''}%
\providecommand \bibnamefont  [1]{#1}%
\providecommand \bibfnamefont [1]{#1}%
\providecommand \citenamefont [1]{#1}%
\providecommand \href@noop [0]{\@secondoftwo}%
\providecommand \href [0]{\begingroup \@sanitize@url \@href}%
\providecommand \@href[1]{\@@startlink{#1}\@@href}%
\providecommand \@@href[1]{\endgroup#1\@@endlink}%
\providecommand \@sanitize@url [0]{\catcode `\\12\catcode `\$12\catcode `\&12\catcode `\#12\catcode `\^12\catcode `\_12\catcode `\%12\relax}%
\providecommand \@@startlink[1]{}%
\providecommand \@@endlink[0]{}%
\providecommand \url  [0]{\begingroup\@sanitize@url \@url }%
\providecommand \@url [1]{\endgroup\@href {#1}{\urlprefix }}%
\providecommand \urlprefix  [0]{URL }%
\providecommand \Eprint [0]{\href }%
\providecommand \doibase [0]{https://doi.org/}%
\providecommand \selectlanguage [0]{\@gobble}%
\providecommand \bibinfo  [0]{\@secondoftwo}%
\providecommand \bibfield  [0]{\@secondoftwo}%
\providecommand \translation [1]{[#1]}%
\providecommand \BibitemOpen [0]{}%
\providecommand \bibitemStop [0]{}%
\providecommand \bibitemNoStop [0]{.\EOS\space}%
\providecommand \EOS [0]{\spacefactor3000\relax}%
\providecommand \BibitemShut  [1]{\csname bibitem#1\endcsname}%
\let\auto@bib@innerbib\@empty
\bibitem [{\citenamefont {Zaman}\ \emph {et~al.}(2011)\citenamefont {Zaman}, \citenamefont {Bridges},\ and\ \citenamefont {Huff}}]{zaman2011evolution}%
  \BibitemOpen
  \bibfield  {author} {\bibinfo {author} {\bibfnamefont {K.~B. M.~Q.}\ \bibnamefont {Zaman}}, \bibinfo {author} {\bibfnamefont {J.~E.}\ \bibnamefont {Bridges}},\ and\ \bibinfo {author} {\bibfnamefont {D.~L.}\ \bibnamefont {Huff}},\ }\bibfield  {title} {\bibinfo {title} {{E}volution from ‘tabs’ to ‘chevron technology’--a review},\ }\href@noop {} {\bibfield  {journal} {\bibinfo  {journal} {International Journal of Aeroacoustics}\ }\textbf {\bibinfo {volume} {10}},\ \bibinfo {pages} {685} (\bibinfo {year} {2011})}\BibitemShut {NoStop}%
\bibitem [{\citenamefont {Gudmundsson}\ and\ \citenamefont {Colonius}(2007)}]{gudmundsson2007spatial}%
  \BibitemOpen
  \bibfield  {author} {\bibinfo {author} {\bibfnamefont {K.}~\bibnamefont {Gudmundsson}}\ and\ \bibinfo {author} {\bibfnamefont {T.}~\bibnamefont {Colonius}},\ }\bibfield  {title} {\bibinfo {title} {{S}patial stability analysis of chevron jet profiles},\ }in\ \href@noop {} {\emph {\bibinfo {booktitle} {13th AIAA/CEAS Aeroacoustics Conference (28th AIAA Aeroacoustics Conference)}}}\ (\bibinfo {year} {2007})\ p.\ \bibinfo {pages} {3599}\BibitemShut {NoStop}%
\bibitem [{\citenamefont {Bridges}\ and\ \citenamefont {Brown}(2004{\natexlab{a}})}]{bridges2004parametric}%
  \BibitemOpen
  \bibfield  {author} {\bibinfo {author} {\bibfnamefont {J.}~\bibnamefont {Bridges}}\ and\ \bibinfo {author} {\bibfnamefont {C.}~\bibnamefont {Brown}},\ }\bibfield  {title} {\bibinfo {title} {{P}arametric testing of chevrons on single flow hot jets},\ }in\ \href@noop {} {\emph {\bibinfo {booktitle} {10th AIAA/CEAS aeroacoustics conference}}}\ (\bibinfo {year} {2004})\ p.\ \bibinfo {pages} {2824}\BibitemShut {NoStop}%
\bibitem [{\citenamefont {Kamliya~Jawahar}\ \emph {et~al.}(2021{\natexlab{a}})\citenamefont {Kamliya~Jawahar}, \citenamefont {Markesteijn}, \citenamefont {Karabasov},\ and\ \citenamefont {Azarpeyvand}}]{kamliya2021effects}%
  \BibitemOpen
  \bibfield  {author} {\bibinfo {author} {\bibfnamefont {H.}~\bibnamefont {Kamliya~Jawahar}}, \bibinfo {author} {\bibfnamefont {A.~P.}\ \bibnamefont {Markesteijn}}, \bibinfo {author} {\bibfnamefont {S.~A.}\ \bibnamefont {Karabasov}},\ and\ \bibinfo {author} {\bibfnamefont {M.}~\bibnamefont {Azarpeyvand}},\ }\bibfield  {title} {\bibinfo {title} {{E}ffects of {C}hevrons on {J}et-installation {N}oise},\ }in\ \href@noop {} {\emph {\bibinfo {booktitle} {AIAA AVIATION 2021 FORUM}}}\ (\bibinfo {year} {2021})\ p.\ \bibinfo {pages} {2184}\BibitemShut {NoStop}%
\bibitem [{\citenamefont {Krasheninnikov}\ and\ \citenamefont {Mironov}(2003{\natexlab{a}})}]{krasheninnikov2003effect}%
  \BibitemOpen
  \bibfield  {author} {\bibinfo {author} {\bibfnamefont {S.~Y.}\ \bibnamefont {Krasheninnikov}}\ and\ \bibinfo {author} {\bibfnamefont {A.~K.}\ \bibnamefont {Mironov}},\ }\bibfield  {title} {\bibinfo {title} {{E}ffect of the streamwise component of the vorticity formed in a turbulent jet source on the acoustic characteristics of the jet},\ }\href@noop {} {\bibfield  {journal} {\bibinfo  {journal} {Fluid Dynamics}\ }\textbf {\bibinfo {volume} {38}},\ \bibinfo {pages} {698} (\bibinfo {year} {2003}{\natexlab{a}})}\BibitemShut {NoStop}%
\bibitem [{\citenamefont {Krasheninnikov}\ and\ \citenamefont {Mironov}(2003{\natexlab{b}})}]{krasheninnikov2003turbulent}%
  \BibitemOpen
  \bibfield  {author} {\bibinfo {author} {\bibfnamefont {S.~Y.}\ \bibnamefont {Krasheninnikov}}\ and\ \bibinfo {author} {\bibfnamefont {A.~K.}\ \bibnamefont {Mironov}},\ }\bibfield  {title} {\bibinfo {title} {{T}urbulent jet acoustic characteristics behavior under additional vorticity creation in jet origin},\ }in\ \href@noop {} {\emph {\bibinfo {booktitle} {Proceedings of the Tenth International Congress on Sound and Vibration}}}\ (\bibinfo {year} {2003})\ pp.\ \bibinfo {pages} {645--652}\BibitemShut {NoStop}%
\bibitem [{\citenamefont {Nikam}\ and\ \citenamefont {Sharma}(2014)}]{nikam2014aero}%
  \BibitemOpen
  \bibfield  {author} {\bibinfo {author} {\bibfnamefont {S.~R.}\ \bibnamefont {Nikam}}\ and\ \bibinfo {author} {\bibfnamefont {S.~D.}\ \bibnamefont {Sharma}},\ }\bibfield  {title} {\bibinfo {title} {{A}ero-acoustic {C}haracteristics of {C}ompressible {J}ets from {C}hevron {N}ozzle},\ }in\ \href@noop {} {\emph {\bibinfo {booktitle} {20th AIAA/CEAS Aeroacoustics Conference}}}\ (\bibinfo {year} {2014})\ p.\ \bibinfo {pages} {2623}\BibitemShut {NoStop}%
\bibitem [{\citenamefont {Saiyed}\ \emph {et~al.}(2000)\citenamefont {Saiyed}, \citenamefont {Bridges},\ and\ \citenamefont {Mikkelsen}}]{saiyed2000acoustics}%
  \BibitemOpen
  \bibfield  {author} {\bibinfo {author} {\bibfnamefont {N.}~\bibnamefont {Saiyed}}, \bibinfo {author} {\bibfnamefont {J.}~\bibnamefont {Bridges}},\ and\ \bibinfo {author} {\bibfnamefont {K.}~\bibnamefont {Mikkelsen}},\ }\bibfield  {title} {\bibinfo {title} {{A}coustics and thrust of separate-flow exhaust nozzles with mixing devices for high-bypass-ratio engines},\ }in\ \href@noop {} {\emph {\bibinfo {booktitle} {6th AIAA/CEAS Aeroacoustics Conference and Exhibit}}}\ (\bibinfo {year} {2000})\ p.\ \bibinfo {pages} {1961}\BibitemShut {NoStop}%
\bibitem [{\citenamefont {Nikam}\ and\ \citenamefont {Sharma}(2017)}]{nikam2017effect}%
  \BibitemOpen
  \bibfield  {author} {\bibinfo {author} {\bibfnamefont {S.~R.}\ \bibnamefont {Nikam}}\ and\ \bibinfo {author} {\bibfnamefont {S.~D.}\ \bibnamefont {Sharma}},\ }\bibfield  {title} {\bibinfo {title} {{E}ffect of chevron nozzle penetration on aero-acoustic characteristics of jet at {M}= 0.8},\ }\href@noop {} {\bibfield  {journal} {\bibinfo  {journal} {Fluid Dynamics Research}\ }\textbf {\bibinfo {volume} {49}},\ \bibinfo {pages} {065506} (\bibinfo {year} {2017})}\BibitemShut {NoStop}%
\bibitem [{\citenamefont {Mengle}\ \emph {et~al.}(2006)\citenamefont {Mengle}, \citenamefont {Elkoby}, \citenamefont {Brusniak},\ and\ \citenamefont {Thomas}}]{mengle2006reducing}%
  \BibitemOpen
  \bibfield  {author} {\bibinfo {author} {\bibfnamefont {V.}~\bibnamefont {Mengle}}, \bibinfo {author} {\bibfnamefont {R.}~\bibnamefont {Elkoby}}, \bibinfo {author} {\bibfnamefont {L.}~\bibnamefont {Brusniak}},\ and\ \bibinfo {author} {\bibfnamefont {R.}~\bibnamefont {Thomas}},\ }\bibfield  {title} {\bibinfo {title} {{R}educing propulsion airframe aeroacoustic interactions with uniquely tailored chevrons: 1. {I}solated nozzles},\ }in\ \href@noop {} {\emph {\bibinfo {booktitle} {12th AIAA/CEAS Aeroacoustics Conference (27th AIAA Aeroacoustics Conference)}}}\ (\bibinfo {year} {2006})\ p.\ \bibinfo {pages} {2467}\BibitemShut {NoStop}%
\bibitem [{\citenamefont {Mead}\ and\ \citenamefont {Strange}(1998)}]{Mead1998}%
  \BibitemOpen
  \bibfield  {author} {\bibinfo {author} {\bibfnamefont {C.}~\bibnamefont {Mead}}\ and\ \bibinfo {author} {\bibfnamefont {P.}~\bibnamefont {Strange}},\ }\bibfield  {title} {\bibinfo {title} {{U}nder-wing installation effects on jet noise at sideline},\ }in\ \href@noop {} {\emph {\bibinfo {booktitle} {4th AIAA/CEAS Aeroacoustics Conference}}}\ (\bibinfo  {publisher} {American Institute of Aeronautics and Astronautics},\ \bibinfo {year} {1998})\ p.\ \bibinfo {pages} {2207}\BibitemShut {NoStop}%
\bibitem [{\citenamefont {Hunter}\ \emph {et~al.}(2005)\citenamefont {Hunter}, \citenamefont {Thomas}, \citenamefont {Abdol-Hamid}, \citenamefont {Pao}, \citenamefont {Elmiligui},\ and\ \citenamefont {Massey}}]{hunter2005computational}%
  \BibitemOpen
  \bibfield  {author} {\bibinfo {author} {\bibfnamefont {C.}~\bibnamefont {Hunter}}, \bibinfo {author} {\bibfnamefont {R.}~\bibnamefont {Thomas}}, \bibinfo {author} {\bibfnamefont {K.}~\bibnamefont {Abdol-Hamid}}, \bibinfo {author} {\bibfnamefont {S.~P.}\ \bibnamefont {Pao}}, \bibinfo {author} {\bibfnamefont {A.}~\bibnamefont {Elmiligui}},\ and\ \bibinfo {author} {\bibfnamefont {S.}~\bibnamefont {Massey}},\ }\bibfield  {title} {\bibinfo {title} {{C}omputational analysis of the flow and acoustic effects of jet-pylon interaction},\ }in\ \href@noop {} {\emph {\bibinfo {booktitle} {11th AIAA/CEAS Aeroacoustics Conference}}}\ (\bibinfo {year} {2005})\ p.\ \bibinfo {pages} {3083}\BibitemShut {NoStop}%
\bibitem [{\citenamefont {Elkoby}(2005)}]{elkoby2005full}%
  \BibitemOpen
  \bibfield  {author} {\bibinfo {author} {\bibfnamefont {R.}~\bibnamefont {Elkoby}},\ }\bibfield  {title} {\bibinfo {title} {{F}ull-scale propulsion airframe aeroacoustics investigation},\ }in\ \href@noop {} {\emph {\bibinfo {booktitle} {11th AIAA/CEAS Aeroacoustics Conference}}}\ (\bibinfo {year} {2005})\ p.\ \bibinfo {pages} {2807}\BibitemShut {NoStop}%
\bibitem [{\citenamefont {Cavalieri}\ \emph {et~al.}(2014)\citenamefont {Cavalieri}, \citenamefont {Jordan}, \citenamefont {Wolf},\ and\ \citenamefont {Gervais}}]{Cavalieri2014}%
  \BibitemOpen
  \bibfield  {author} {\bibinfo {author} {\bibfnamefont {A.~V.~G.}\ \bibnamefont {Cavalieri}}, \bibinfo {author} {\bibfnamefont {P.}~\bibnamefont {Jordan}}, \bibinfo {author} {\bibfnamefont {W.}~\bibnamefont {Wolf}},\ and\ \bibinfo {author} {\bibfnamefont {Y.}~\bibnamefont {Gervais}},\ }\bibfield  {title} {\bibinfo {title} {{S}cattering of wavepackets by a flat plate in the vicinity of a turbulent jet},\ }\href@noop {} {\bibfield  {journal} {\bibinfo  {journal} {Journal of Sound and Vibration}\ }\textbf {\bibinfo {volume} {333}},\ \bibinfo {pages} {6516} (\bibinfo {year} {2014})}\BibitemShut {NoStop}%
\bibitem [{\citenamefont {Lyu}\ \emph {et~al.}(2017)\citenamefont {Lyu}, \citenamefont {Dowling},\ and\ \citenamefont {Naqavi}}]{Lyu2017a}%
  \BibitemOpen
  \bibfield  {author} {\bibinfo {author} {\bibfnamefont {B.}~\bibnamefont {Lyu}}, \bibinfo {author} {\bibfnamefont {A.~P.}\ \bibnamefont {Dowling}},\ and\ \bibinfo {author} {\bibfnamefont {I.}~\bibnamefont {Naqavi}},\ }\bibfield  {title} {\bibinfo {title} {{P}rediction of installed jet noise},\ }\href@noop {} {\bibfield  {journal} {\bibinfo  {journal} {Journal of Fluid Mechanics}\ }\textbf {\bibinfo {volume} {811}},\ \bibinfo {pages} {234} (\bibinfo {year} {2017})}\BibitemShut {NoStop}%
\bibitem [{\citenamefont {Mayoral}\ and\ \citenamefont {Papamoschou}(2010)}]{mayoral2010effects}%
  \BibitemOpen
  \bibfield  {author} {\bibinfo {author} {\bibfnamefont {S.}~\bibnamefont {Mayoral}}\ and\ \bibinfo {author} {\bibfnamefont {D.}~\bibnamefont {Papamoschou}},\ }\bibfield  {title} {\bibinfo {title} {{E}ffects of source redistribution on jet noise shielding},\ }in\ \href@noop {} {\emph {\bibinfo {booktitle} {48th AIAA Aerospace Sciences Meeting Including the New Horizons Forum and Aerospace Exposition}}}\ (\bibinfo {year} {2010})\ p.\ \bibinfo {pages} {652}\BibitemShut {NoStop}%
\bibitem [{\citenamefont {Mengle}(2011)}]{mengle2011effect}%
  \BibitemOpen
  \bibfield  {author} {\bibinfo {author} {\bibfnamefont {V.}~\bibnamefont {Mengle}},\ }\bibfield  {title} {\bibinfo {title} {{T}he effect of nozzle-to-wing gulley height on jet flow attachment to the wing and jet-flap interaction noise},\ }in\ \href@noop {} {\emph {\bibinfo {booktitle} {17th AIAA/CEAS Aeroacoustics Conference (32nd AIAA Aeroacoustics Conference)}}}\ (\bibinfo {year} {2011})\ p.\ \bibinfo {pages} {2705}\BibitemShut {NoStop}%
\bibitem [{\citenamefont {Mengle}\ \emph {et~al.}(2007)\citenamefont {Mengle}, \citenamefont {Stoker}, \citenamefont {Brusniak}, \citenamefont {Elkoby},\ and\ \citenamefont {Thomas}}]{mengle2007flaperon}%
  \BibitemOpen
  \bibfield  {author} {\bibinfo {author} {\bibfnamefont {V.}~\bibnamefont {Mengle}}, \bibinfo {author} {\bibfnamefont {R.}~\bibnamefont {Stoker}}, \bibinfo {author} {\bibfnamefont {L.}~\bibnamefont {Brusniak}}, \bibinfo {author} {\bibfnamefont {R.}~\bibnamefont {Elkoby}},\ and\ \bibinfo {author} {\bibfnamefont {R.}~\bibnamefont {Thomas}},\ }\bibfield  {title} {\bibinfo {title} {{F}laperon modification effect on jet-flap interaction noise reduction for chevron nozzles},\ }in\ \href@noop {} {\emph {\bibinfo {booktitle} {13th AIAA/CEAS Aeroacoustics Conference (28th AIAA Aeroacoustics Conference)}}}\ (\bibinfo {year} {2007})\ p.\ \bibinfo {pages} {3666}\BibitemShut {NoStop}%
\bibitem [{\citenamefont {Kopiev}\ \emph {et~al.}(2013)\citenamefont {Kopiev}, \citenamefont {Faranosov}, \citenamefont {Zaitsev}, \citenamefont {Vlasov}, \citenamefont {Belyaev}, \citenamefont {Ostrikov},\ and\ \citenamefont {Karavosov}}]{kopiev2013intensification}%
  \BibitemOpen
  \bibfield  {author} {\bibinfo {author} {\bibfnamefont {V.~F.}\ \bibnamefont {Kopiev}}, \bibinfo {author} {\bibfnamefont {G.~A.}\ \bibnamefont {Faranosov}}, \bibinfo {author} {\bibfnamefont {M.}~\bibnamefont {Zaitsev}}, \bibinfo {author} {\bibfnamefont {E.}~\bibnamefont {Vlasov}}, \bibinfo {author} {\bibfnamefont {I.}~\bibnamefont {Belyaev}}, \bibinfo {author} {\bibfnamefont {N.}~\bibnamefont {Ostrikov}},\ and\ \bibinfo {author} {\bibfnamefont {R.}~\bibnamefont {Karavosov}},\ }\bibfield  {title} {\bibinfo {title} {{I}ntensification and suppression of jet noise sources in the vicinity of lifting surfaces},\ }in\ \href@noop {} {\emph {\bibinfo {booktitle} {19th AIAA/CEAS Aeroacoustics Conference}}}\ (\bibinfo {year} {2013})\ p.\ \bibinfo {pages} {2284}\BibitemShut {NoStop}%
\bibitem [{\citenamefont {Bastos}\ \emph {et~al.}(2017)\citenamefont {Bastos}, \citenamefont {Deschamps},\ and\ \citenamefont {da~Silva}}]{bastos2017experimental}%
  \BibitemOpen
  \bibfield  {author} {\bibinfo {author} {\bibfnamefont {L.~P.}\ \bibnamefont {Bastos}}, \bibinfo {author} {\bibfnamefont {C.~J.}\ \bibnamefont {Deschamps}},\ and\ \bibinfo {author} {\bibfnamefont {A.~R.}\ \bibnamefont {da~Silva}},\ }\bibfield  {title} {\bibinfo {title} {{E}xperimental investigation of the far-field noise due to jet-surface interaction combined with a chevron nozzle},\ }\href@noop {} {\bibfield  {journal} {\bibinfo  {journal} {Applied Acoustics}\ }\textbf {\bibinfo {volume} {127}},\ \bibinfo {pages} {240} (\bibinfo {year} {2017})}\BibitemShut {NoStop}%
\bibitem [{\citenamefont {Kamliya~Jawahar}\ and\ \citenamefont {Azarpeyvand}(2022)}]{kamliya2022Hydro}%
  \BibitemOpen
  \bibfield  {author} {\bibinfo {author} {\bibfnamefont {H.}~\bibnamefont {Kamliya~Jawahar}}\ and\ \bibinfo {author} {\bibfnamefont {M.}~\bibnamefont {Azarpeyvand}},\ }\bibfield  {title} {\bibinfo {title} {{O}n {I}nvestigating the {H}ydrodynamic {F}ield for {J}ets with and without {I}nstallation {E}ffects},\ }in\ \href@noop {} {\emph {\bibinfo {booktitle} {AIAA AVIATION 2022 FORUM}}}\ (\bibinfo {year} {2022})\ p.\ \bibinfo {pages} {3791}\BibitemShut {NoStop}%
\bibitem [{\citenamefont {Carbini}\ \emph {et~al.}(2022)\citenamefont {Carbini}, \citenamefont {Meloni}, \citenamefont {Camussi}, \citenamefont {Lawrence},\ and\ \citenamefont {Ramos~Proenca}}]{carbini2022experimental}%
  \BibitemOpen
  \bibfield  {author} {\bibinfo {author} {\bibfnamefont {E.}~\bibnamefont {Carbini}}, \bibinfo {author} {\bibfnamefont {S.}~\bibnamefont {Meloni}}, \bibinfo {author} {\bibfnamefont {R.}~\bibnamefont {Camussi}}, \bibinfo {author} {\bibfnamefont {J.}~\bibnamefont {Lawrence}},\ and\ \bibinfo {author} {\bibfnamefont {A.}~\bibnamefont {Ramos~Proenca}},\ }\bibfield  {title} {\bibinfo {title} {{A}n experimental investigation into the influence of installed chevron jet flows on wall-pressure fluctuations},\ }in\ \href@noop {} {\emph {\bibinfo {booktitle} {Proceedings of the Institute of Acoustics}}},\ Vol.~\bibinfo {volume} {44}\ (\bibinfo  {publisher} {Institute of Acoustics},\ \bibinfo {year} {2022})\BibitemShut {NoStop}%
\bibitem [{\citenamefont {Carbini}\ \emph {et~al.}(2023)\citenamefont {Carbini}, \citenamefont {Lawrence}, \citenamefont {Proen{\c{c}}a}, \citenamefont {Meloni},\ and\ \citenamefont {Camussi}}]{carbini2023experimental}%
  \BibitemOpen
  \bibfield  {author} {\bibinfo {author} {\bibfnamefont {E.}~\bibnamefont {Carbini}}, \bibinfo {author} {\bibfnamefont {J.}~\bibnamefont {Lawrence}}, \bibinfo {author} {\bibfnamefont {A.}~\bibnamefont {Proen{\c{c}}a}}, \bibinfo {author} {\bibfnamefont {S.}~\bibnamefont {Meloni}},\ and\ \bibinfo {author} {\bibfnamefont {R.}~\bibnamefont {Camussi}},\ }\bibfield  {title} {\bibinfo {title} {{A}n experimental investigation into the influence of installed passively controlled jet flows on wall-pressure fluctuations},\ }in\ \href@noop {} {\emph {\bibinfo {booktitle} {INTER-NOISE and NOISE-CON Congress and Conference Proceedings}}},\ Vol.\ \bibinfo {volume} {265}\ (\bibinfo {year} {2023})\ pp.\ \bibinfo {pages} {2350--2357}\BibitemShut {NoStop}%
\bibitem [{\citenamefont {Lyu}\ and\ \citenamefont {Dowling}(2017)}]{Lyu2017b}%
  \BibitemOpen
  \bibfield  {author} {\bibinfo {author} {\bibfnamefont {B.}~\bibnamefont {Lyu}}\ and\ \bibinfo {author} {\bibfnamefont {A.~P.}\ \bibnamefont {Dowling}},\ }\bibfield  {title} {\bibinfo {title} {{O}n the mechanism and reduction of installed jet noise},\ }in\ \href@noop {} {\emph {\bibinfo {booktitle} {23rd AIAA/CEAS Aeroacoustics Conference}}}\ (\bibinfo {year} {2017})\ p.\ \bibinfo {pages} {3523}\BibitemShut {NoStop}%
\bibitem [{\citenamefont {Lyu}\ and\ \citenamefont {Dowling}(2019{\natexlab{a}})}]{lyu2019experimental}%
  \BibitemOpen
  \bibfield  {author} {\bibinfo {author} {\bibfnamefont {B.}~\bibnamefont {Lyu}}\ and\ \bibinfo {author} {\bibfnamefont {A.~P.}\ \bibnamefont {Dowling}},\ }\bibfield  {title} {\bibinfo {title} {{A}n experimental study of the effects of lobed nozzles on installed jet noise},\ }\href@noop {} {\bibfield  {journal} {\bibinfo  {journal} {Experiments in Fluids}\ }\textbf {\bibinfo {volume} {60}},\ \bibinfo {pages} {1} (\bibinfo {year} {2019}{\natexlab{a}})}\BibitemShut {NoStop}%
\bibitem [{\citenamefont {Thomas}\ \emph {et~al.}(2001)\citenamefont {Thomas}, \citenamefont {Kinzie},\ and\ \citenamefont {Pao}}]{thomas2001computational}%
  \BibitemOpen
  \bibfield  {author} {\bibinfo {author} {\bibfnamefont {R.~H.}\ \bibnamefont {Thomas}}, \bibinfo {author} {\bibfnamefont {K.~W.}\ \bibnamefont {Kinzie}},\ and\ \bibinfo {author} {\bibfnamefont {S.~P.}\ \bibnamefont {Pao}},\ }\bibfield  {title} {\bibinfo {title} {{C}omputational analysis of a pylon-chevron core nozzle interaction},\ }in\ \href@noop {} {\emph {\bibinfo {booktitle} {7th AIAA/CEAS Aeroacoustics Conference and Exhibit}}}\ (\bibinfo {year} {2001})\ p.\ \bibinfo {pages} {2185}\BibitemShut {NoStop}%
\bibitem [{\citenamefont {Wang}\ \emph {et~al.}(2017)\citenamefont {Wang}, \citenamefont {Tyacke},\ and\ \citenamefont {Tucker}}]{wang2017rans}%
  \BibitemOpen
  \bibfield  {author} {\bibinfo {author} {\bibfnamefont {Z.-N.}\ \bibnamefont {Wang}}, \bibinfo {author} {\bibfnamefont {J.~C.}\ \bibnamefont {Tyacke}},\ and\ \bibinfo {author} {\bibfnamefont {P.~G.}\ \bibnamefont {Tucker}},\ }\bibfield  {title} {\bibinfo {title} {{LES-RANS} of installed ultra-high bypass-ratio coaxial jet aeroacoustics with a finite span wing-flap geometry and flight stream-{P}art 2: chevron nozzle},\ }in\ \href@noop {} {\emph {\bibinfo {booktitle} {23rd AIAA/CEAS Aeroacoustics Conference}}}\ (\bibinfo {year} {2017})\ p.\ \bibinfo {pages} {3855}\BibitemShut {NoStop}%
\bibitem [{\citenamefont {Abid}\ \emph {et~al.}(2022)\citenamefont {Abid}, \citenamefont {Markesteijn}, \citenamefont {Gryazev}, \citenamefont {Karabasov}, \citenamefont {Kamliya~Jawahar},\ and\ \citenamefont {Azarpeyvand}}]{abid2022jet}%
  \BibitemOpen
  \bibfield  {author} {\bibinfo {author} {\bibfnamefont {H.~A.}\ \bibnamefont {Abid}}, \bibinfo {author} {\bibfnamefont {A.~P.}\ \bibnamefont {Markesteijn}}, \bibinfo {author} {\bibfnamefont {V.}~\bibnamefont {Gryazev}}, \bibinfo {author} {\bibfnamefont {S.~A.}\ \bibnamefont {Karabasov}}, \bibinfo {author} {\bibfnamefont {H.}~\bibnamefont {Kamliya~Jawahar}},\ and\ \bibinfo {author} {\bibfnamefont {M.}~\bibnamefont {Azarpeyvand}},\ }\bibfield  {title} {\bibinfo {title} {{J}et {I}nstallation {N}oise {M}odelling {I}nformed by {GPU} {LES}},\ }in\ \href@noop {} {\emph {\bibinfo {booktitle} {28th AIAA/CEAS Aeroacoustics 2022 Conference}}}\ (\bibinfo {year} {2022})\ p.\ \bibinfo {pages} {2906}\BibitemShut {NoStop}%
\bibitem [{\citenamefont {Abid}\ \emph {et~al.}(2023)\citenamefont {Abid}, \citenamefont {Markesteijn}, \citenamefont {Gryazev}, \citenamefont {Toropov}, \citenamefont {Karabasov}, \citenamefont {Yang}, \citenamefont {Allen}, \citenamefont {Kamliya~Jawahar},\ and\ \citenamefont {Azarpeyvand}}]{abid2023influence}%
  \BibitemOpen
  \bibfield  {author} {\bibinfo {author} {\bibfnamefont {H.~A.}\ \bibnamefont {Abid}}, \bibinfo {author} {\bibfnamefont {A.~P.}\ \bibnamefont {Markesteijn}}, \bibinfo {author} {\bibfnamefont {V.}~\bibnamefont {Gryazev}}, \bibinfo {author} {\bibfnamefont {V.}~\bibnamefont {Toropov}}, \bibinfo {author} {\bibfnamefont {S.~A.}\ \bibnamefont {Karabasov}}, \bibinfo {author} {\bibfnamefont {G.}~\bibnamefont {Yang}}, \bibinfo {author} {\bibfnamefont {C.~B.}\ \bibnamefont {Allen}}, \bibinfo {author} {\bibfnamefont {H.}~\bibnamefont {Kamliya~Jawahar}},\ and\ \bibinfo {author} {\bibfnamefont {M.}~\bibnamefont {Azarpeyvand}},\ }\bibfield  {title} {\bibinfo {title} {{I}nfluence of {N}ozzle {G}eometry on {J}et mixing and {J}et {I}nstallation {N}oise},\ }in\ \href@noop {} {\emph {\bibinfo {booktitle} {AIAA SCITECH 2023 Forum}}}\ (\bibinfo {year} {2023})\ p.\ \bibinfo {pages} {0615}\BibitemShut {NoStop}%
\bibitem [{\citenamefont {Lyu}\ and\ \citenamefont {Dowling}(2018{\natexlab{a}})}]{Lyu2018a}%
  \BibitemOpen
  \bibfield  {author} {\bibinfo {author} {\bibfnamefont {B.}~\bibnamefont {Lyu}}\ and\ \bibinfo {author} {\bibfnamefont {A.~P.}\ \bibnamefont {Dowling}},\ }\bibfield  {title} {\bibinfo {title} {{E}xperimental validation of the hybrid scattering model for installed jet noise},\ }\href@noop {} {\bibfield  {journal} {\bibinfo  {journal} {Physics of Fluids}\ }\textbf {\bibinfo {volume} {30}},\ \bibinfo {pages} {085102} (\bibinfo {year} {2018}{\natexlab{a}})}\BibitemShut {NoStop}%
\bibitem [{\citenamefont {Kamliya~Jawahar}\ \emph {et~al.}(2021{\natexlab{b}})\citenamefont {Kamliya~Jawahar}, \citenamefont {Baskaran},\ and\ \citenamefont {Azarpeyvand}}]{Screech2021}%
  \BibitemOpen
  \bibfield  {author} {\bibinfo {author} {\bibfnamefont {H.}~\bibnamefont {Kamliya~Jawahar}}, \bibinfo {author} {\bibfnamefont {K.}~\bibnamefont {Baskaran}},\ and\ \bibinfo {author} {\bibfnamefont {M.}~\bibnamefont {Azarpeyvand}},\ }\bibfield  {title} {\bibinfo {title} {{U}nsteady {C}haracteristics of {M}ode {O}scillation for {S}creeching {J}ets},\ }in\ \href@noop {} {\emph {\bibinfo {booktitle} {AIAA AVIATION 2021 FORUM}}}\ (\bibinfo {year} {2021})\ p.\ \bibinfo {pages} {2279}\BibitemShut {NoStop}%
\bibitem [{\citenamefont {Kamliya~Jawahar}\ \emph {et~al.}(2021{\natexlab{c}})\citenamefont {Kamliya~Jawahar}, \citenamefont {Meloni}, \citenamefont {Camussi},\ and\ \citenamefont {Azarpeyvand}}]{ChevronMach2021}%
  \BibitemOpen
  \bibfield  {author} {\bibinfo {author} {\bibfnamefont {H.}~\bibnamefont {Kamliya~Jawahar}}, \bibinfo {author} {\bibfnamefont {S.}~\bibnamefont {Meloni}}, \bibinfo {author} {\bibfnamefont {R.}~\bibnamefont {Camussi}},\ and\ \bibinfo {author} {\bibfnamefont {M.}~\bibnamefont {Azarpeyvand}},\ }\bibfield  {title} {\bibinfo {title} {{E}xperimental {I}nvestigation on the {J}et {N}oise {S}ources for {C}hevron {N}ozzles in {U}nder-expanded {C}ondition},\ }in\ \href@noop {} {\emph {\bibinfo {booktitle} {AIAA AVIATION 2021 FORUM}}}\ (\bibinfo {year} {2021})\ p.\ \bibinfo {pages} {2181}\BibitemShut {NoStop}%
\bibitem [{\citenamefont {Kamliya~Jawahar}\ and\ \citenamefont {Azarpeyvand}(2021)}]{Porous2021}%
  \BibitemOpen
  \bibfield  {author} {\bibinfo {author} {\bibfnamefont {H.}~\bibnamefont {Kamliya~Jawahar}}\ and\ \bibinfo {author} {\bibfnamefont {M.}~\bibnamefont {Azarpeyvand}},\ }\bibfield  {title} {\bibinfo {title} {{T}railing-edge {T}reatments for {J}et-installation {N}oise {R}eduction},\ }in\ \href@noop {} {\emph {\bibinfo {booktitle} {AIAA AVIATION 2021 FORUM}}}\ (\bibinfo {year} {2021})\ p.\ \bibinfo {pages} {2185}\BibitemShut {NoStop}%
\bibitem [{\citenamefont {Kamliya~Jawahar}\ \emph {et~al.}(2021{\natexlab{d}})\citenamefont {Kamliya~Jawahar}, \citenamefont {Markesteijn}, \citenamefont {Karabasov},\ and\ \citenamefont {Azarpeyvand}}]{Chevron2021}%
  \BibitemOpen
  \bibfield  {author} {\bibinfo {author} {\bibfnamefont {H.}~\bibnamefont {Kamliya~Jawahar}}, \bibinfo {author} {\bibfnamefont {A.~P.}\ \bibnamefont {Markesteijn}}, \bibinfo {author} {\bibfnamefont {S.~A.}\ \bibnamefont {Karabasov}},\ and\ \bibinfo {author} {\bibfnamefont {M.}~\bibnamefont {Azarpeyvand}},\ }\bibfield  {title} {\bibinfo {title} {{E}ffects of {C}hevrons on {J}et-installation {N}oise},\ }in\ \href@noop {} {\emph {\bibinfo {booktitle} {AIAA AVIATION 2021 FORUM}}}\ (\bibinfo {year} {2021})\ p.\ \bibinfo {pages} {2184}\BibitemShut {NoStop}%
\bibitem [{\citenamefont {Mayer}\ \emph {et~al.}(2019)\citenamefont {Mayer}, \citenamefont {Jawahar}, \citenamefont {Sz{\H{o}}ke}, \citenamefont {Ali},\ and\ \citenamefont {Azarpeyvand}}]{Mayer2019}%
  \BibitemOpen
  \bibfield  {author} {\bibinfo {author} {\bibfnamefont {Y.~D.}\ \bibnamefont {Mayer}}, \bibinfo {author} {\bibfnamefont {H.~K.}\ \bibnamefont {Jawahar}}, \bibinfo {author} {\bibfnamefont {M.}~\bibnamefont {Sz{\H{o}}ke}}, \bibinfo {author} {\bibfnamefont {S.~A.~S.}\ \bibnamefont {Ali}},\ and\ \bibinfo {author} {\bibfnamefont {M.}~\bibnamefont {Azarpeyvand}},\ }\bibfield  {title} {\bibinfo {title} {{D}esign and performance of an aeroacoustic wind tunnel facility at the {U}niversity of {B}ristol},\ }\href@noop {} {\bibfield  {journal} {\bibinfo  {journal} {Applied Acoustics}\ }\textbf {\bibinfo {volume} {155}},\ \bibinfo {pages} {358} (\bibinfo {year} {2019})}\BibitemShut {NoStop}%
\bibitem [{\citenamefont {Bridges}\ and\ \citenamefont {Brown}(2004{\natexlab{b}})}]{NASAChevron}%
  \BibitemOpen
  \bibfield  {author} {\bibinfo {author} {\bibfnamefont {J.}~\bibnamefont {Bridges}}\ and\ \bibinfo {author} {\bibfnamefont {C.}~\bibnamefont {Brown}},\ }\bibfield  {title} {\bibinfo {title} {{P}arametric testing of chevrons on single flow hot jets},\ }in\ \href@noop {} {\emph {\bibinfo {booktitle} {10th AIAA/CEAS Aeroacoustics Conference}}}\ (\bibinfo {year} {2004})\ p.\ \bibinfo {pages} {2824}\BibitemShut {NoStop}%
\bibitem [{\citenamefont {Lyu}\ and\ \citenamefont {Dowling}(2018{\natexlab{b}})}]{Lyu2018h}%
  \BibitemOpen
  \bibfield  {author} {\bibinfo {author} {\bibfnamefont {B.}~\bibnamefont {Lyu}}\ and\ \bibinfo {author} {\bibfnamefont {A.~P.}\ \bibnamefont {Dowling}},\ }\bibfield  {title} {\bibinfo {title} {{P}rediction of installed jet noise due to swept wings},\ }in\ \href@noop {} {\emph {\bibinfo {booktitle} {24th AIAA/CEAS Aeroacoustics Conference}}}\ (\bibinfo {year} {2018})\ p.\ \bibinfo {pages} {2980}\BibitemShut {NoStop}%
\bibitem [{\citenamefont {Lyu}\ and\ \citenamefont {Dowling}(2019{\natexlab{b}})}]{Lyu2019a}%
  \BibitemOpen
  \bibfield  {author} {\bibinfo {author} {\bibfnamefont {B.}~\bibnamefont {Lyu}}\ and\ \bibinfo {author} {\bibfnamefont {A.~P.}\ \bibnamefont {Dowling}},\ }\bibfield  {title} {\bibinfo {title} {{M}odelling installed jet noise due to the scattering of jet instability waves by swept wings},\ }\href@noop {} {\bibfield  {journal} {\bibinfo  {journal} {Journal of Fluid Mechanics}\ }\textbf {\bibinfo {volume} {870}},\ \bibinfo {pages} {760} (\bibinfo {year} {2019}{\natexlab{b}})}\BibitemShut {NoStop}%
\bibitem [{\citenamefont {Viswanathan}(2003)}]{viswanathan2003jet}%
  \BibitemOpen
  \bibfield  {author} {\bibinfo {author} {\bibfnamefont {K.}~\bibnamefont {Viswanathan}},\ }\bibfield  {title} {\bibinfo {title} {{J}et aeroacoustic testing: issues and implications},\ }\href@noop {} {\bibfield  {journal} {\bibinfo  {journal} {AIAA Journal}\ }\textbf {\bibinfo {volume} {41}},\ \bibinfo {pages} {1674} (\bibinfo {year} {2003})}\BibitemShut {NoStop}%
\bibitem [{\citenamefont {Viswanathan}(2006)}]{viswanathan2006instrumentation}%
  \BibitemOpen
  \bibfield  {author} {\bibinfo {author} {\bibfnamefont {K.}~\bibnamefont {Viswanathan}},\ }\bibfield  {title} {\bibinfo {title} {{I}nstrumentation considerations for accurate jet noise measurements},\ }\href@noop {} {\bibfield  {journal} {\bibinfo  {journal} {AIAA Journal}\ }\textbf {\bibinfo {volume} {44}},\ \bibinfo {pages} {1137} (\bibinfo {year} {2006})}\BibitemShut {NoStop}%
\bibitem [{\citenamefont {Brown}\ and\ \citenamefont {Bridges}(2006)}]{brown2006small}%
  \BibitemOpen
  \bibfield  {author} {\bibinfo {author} {\bibfnamefont {C.}~\bibnamefont {Brown}}\ and\ \bibinfo {author} {\bibfnamefont {J.}~\bibnamefont {Bridges}},\ }\href@noop {} {\emph {\bibinfo {title} {{S}mall hot jet acoustic rig validation}}},\ \bibinfo {type} {Tech. Rep.}\ \bibinfo {number} {NASA/TM-2006-214413}\ (\bibinfo  {institution} {NASA Glenn Research Center},\ \bibinfo {year} {2006})\BibitemShut {NoStop}%
\bibitem [{\citenamefont {Bastos}\ \emph {et~al.}(2018)\citenamefont {Bastos}, \citenamefont {Deschamps}, \citenamefont {da~Silva}, \citenamefont {Cordioli}, \citenamefont {Sirotto}, \citenamefont {Maia}, \citenamefont {Coelho},\ and\ \citenamefont {Queiroz}}]{bastos2018development}%
  \BibitemOpen
  \bibfield  {author} {\bibinfo {author} {\bibfnamefont {L.~P.}\ \bibnamefont {Bastos}}, \bibinfo {author} {\bibfnamefont {C.~J.}\ \bibnamefont {Deschamps}}, \bibinfo {author} {\bibfnamefont {A.~R.}\ \bibnamefont {da~Silva}}, \bibinfo {author} {\bibfnamefont {J.~A.}\ \bibnamefont {Cordioli}}, \bibinfo {author} {\bibfnamefont {J.~R.}\ \bibnamefont {Sirotto}}, \bibinfo {author} {\bibfnamefont {I.~A.}\ \bibnamefont {Maia}}, \bibinfo {author} {\bibfnamefont {E.~L.}\ \bibnamefont {Coelho}},\ and\ \bibinfo {author} {\bibfnamefont {R.~L.}\ \bibnamefont {Queiroz}},\ }\bibfield  {title} {\bibinfo {title} {{D}evelopment, validation and application of a newly developed rig facility for investigation of jet aeroacoustics},\ }\href@noop {} {\bibfield  {journal} {\bibinfo  {journal} {Journal of the Brazilian Society of Mechanical Sciences and Engineering}\ }\textbf {\bibinfo {volume} {40}},\ \bibinfo {pages} {1} (\bibinfo {year} {2018})}\BibitemShut {NoStop}%
\bibitem [{\citenamefont {Head}\ and\ \citenamefont {Fisher}(1976)}]{Head1976}%
  \BibitemOpen
  \bibfield  {author} {\bibinfo {author} {\bibfnamefont {R.}~\bibnamefont {Head}}\ and\ \bibinfo {author} {\bibfnamefont {M.}~\bibnamefont {Fisher}},\ }\bibfield  {title} {\bibinfo {title} {{J}et/surface interaction noise - {A}nalysis of farfield low frequency augmentations of jet noise due to the presence of a solid shield},\ }in\ \href@noop {} {\emph {\bibinfo {booktitle} {3rd Aeroacoustics Conference}}}\ (\bibinfo  {publisher} {American Institute of Aeronautics and Astronautics},\ \bibinfo {year} {1976})\ p.\ \bibinfo {pages} {502}\BibitemShut {NoStop}%
\bibitem [{\citenamefont {Jawahar}\ \emph {et~al.}(2023)\citenamefont {Jawahar}, \citenamefont {Karabasov},\ and\ \citenamefont {Azarpeyvand}}]{Porous2023}%
  \BibitemOpen
  \bibfield  {author} {\bibinfo {author} {\bibfnamefont {H.~K.}\ \bibnamefont {Jawahar}}, \bibinfo {author} {\bibfnamefont {S.~A.}\ \bibnamefont {Karabasov}},\ and\ \bibinfo {author} {\bibfnamefont {M.}~\bibnamefont {Azarpeyvand}},\ }\bibfield  {title} {\bibinfo {title} {{J}et installation noise reduction using porous treatments},\ }\href@noop {} {\bibfield  {journal} {\bibinfo  {journal} {Journal of Sound and Vibration}\ }\textbf {\bibinfo {volume} {545}},\ \bibinfo {pages} {117406} (\bibinfo {year} {2023})}\BibitemShut {NoStop}%
\bibitem [{\citenamefont {Suzuki}\ and\ \citenamefont {Colonius}(2006)}]{suzuki2006}%
  \BibitemOpen
  \bibfield  {author} {\bibinfo {author} {\bibfnamefont {T.}~\bibnamefont {Suzuki}}\ and\ \bibinfo {author} {\bibfnamefont {T.}~\bibnamefont {Colonius}},\ }\bibfield  {title} {\bibinfo {title} {{I}nstability waves in a subsonic round jet detected using a near-field phased microphone array},\ }\href@noop {} {\bibfield  {journal} {\bibinfo  {journal} {Journal of Fluid Mechanics}\ }\textbf {\bibinfo {volume} {565}},\ \bibinfo {pages} {197} (\bibinfo {year} {2006})}\BibitemShut {NoStop}%
\bibitem [{\citenamefont {Jordan}\ and\ \citenamefont {Colonius}(2013)}]{Jordan2013}%
  \BibitemOpen
  \bibfield  {author} {\bibinfo {author} {\bibfnamefont {P.}~\bibnamefont {Jordan}}\ and\ \bibinfo {author} {\bibfnamefont {T.}~\bibnamefont {Colonius}},\ }\bibfield  {title} {\bibinfo {title} {{W}ave packets and turbulent jet noise},\ }\href@noop {} {\bibfield  {journal} {\bibinfo  {journal} {Annual Review of Fluid Mechanics}\ }\textbf {\bibinfo {volume} {45}},\ \bibinfo {pages} {173} (\bibinfo {year} {2013})}\BibitemShut {NoStop}%
\end{thebibliography}%

\end{document}